\documentclass[linenumbers]{aastex631}

\usepackage{natbib}
\usepackage{graphicx}
\usepackage{float}
\usepackage{amsmath}
\usepackage{textcomp}

\usepackage{hyperref}
\usepackage[caption=false]{subfig}
\let\splitbox\undefined
\usepackage[export]{adjustbox}

\submitjournal{ApJ}

\shorttitle{Strongly Lensed SMBHBs as nHz GW Sources}
\shortauthors{Khusid, Mingarelli, Natarajan et al.}

\begin{document}

\nolinenumbers

\title{Strongly Lensed Supermassive Black Hole Binaries as Nanohertz Gravitational-Wave Sources}

\correspondingauthor{Nicole Khusid}
\email{nicole.khusid@stonybrook.edu}

\author[0000-0001-9304-7075]{Nicole M. Khusid}
\affiliation{Department of Physics, University of Connecticut, 196 Auditorium Road, U-3046, Storrs, CT 06269-3046, USA}
\affiliation{Department of Physics and Astronomy, Stony Brook University, Stony Brook NY 11794, USA}

\author[0000-0002-4307-1322]{Chiara M. F. Mingarelli}
\affiliation{Department of Physics, University of Connecticut, 196 Auditorium Road, U-3046, Storrs, CT 06269-3046, USA}
\affiliation{Center for Computational Astrophysics, Flatiron Institute, 162 Fifth Ave, New York, NY, 10010, USA}
\affiliation{Department of Physics, Yale University, P. O. Box 208120, New Haven, CT 06520-208120, USA}

\author[0000-0002-5554-8896]{Priyamvada Natarajan}
\affiliation{Department of Astronomy, Yale University, P. O. Box 208101, New Haven, CT 06511-208101, USA}
\affiliation{Department of Physics, Yale University, P. O. Box 208120, New Haven, CT 06520-208120, USA}
\affiliation{Black Hole Initiative, Harvard University, 20 Garden Street, Cambridge, MA 02138, USA}

\author[0000-0002-5557-4007]{J. Andrew Casey-Clyde}
\affiliation{Department of Physics, University of Connecticut, 196 Auditorium Road, U-3046, Storrs, CT 06269-3046, USA}

\author[0000-0001-5655-4158]{Anna Barnacka}
\affiliation{Harvard-Smithsonian Center for Astrophysics, 60 Garden St, MS-20, Cambridge, MA 02138, USA}
\affiliation{Astronomical Observatory, Jagiellonian University, Cracow, Poland}

\begin{abstract}
\nolinenumbers
Supermassive black hole binary systems (SMBHBs) should be the most powerful sources of gravitational waves (GWs) in the Universe. Once Pulsar Timing Arrays (PTAs) detect the stochastic GW background from their cosmic merger history, searching for individually resolvable binaries will take on new importance. Since these individual SMBHBs are expected to be rare, here we explore how strong gravitational lensing can act as a tool for increasing their detection prospects by magnifying fainter sources and bringing them into view. Unlike for electromagnetic waves, when the geometric optics limit is nearly always valid, for GWs the wave-diffraction-interference effects can become important when the wavelength of the GWs is larger than the Schwarzchild radius of the lens, i.e. $M_{\rm lens} \sim 10^8\,(\frac{f}{mHz})^{-1}\,M_\odot$. For the GW frequency range explored in this work, the geometric optics limit holds. We investigate GW signals from SMBHBs that might be detectable with current and future PTAs under the assumption that quasars serve as bright beacons that signal a recent merger. Using the black hole mass function derived from quasars and a physically motivated magnification distribution, we expect to detect a few strongly lensed binary systems out to $z \approx 2$. Additionally, for a range of fixed magnifications $2 \leq \mu \leq 100$, strong lensing adds up to $\sim$30 more detectable binaries for PTAs. Finally, we investigate the possibility of observing both time-delayed electromagnetic signals and GW signals from these strongly lensed binary systems --- that will provide us with unprecedented multimessenger insights into their orbital evolution.\end{abstract}

    \keywords{supermassive black holes -- gravitational waves -- quasars -- pulsars -- strong lensing}

\section{Introduction} \label{sec:intro}

Gravitational waves (GWs) are ripples in the fabric of space-time resulting from e.g. the coalescence of massive compact objects. A tantalizing astrophysical source of GWs which may be detected in the next few years \citep{0.02, Xin} is the inspiral and coalescence of SMBHBs. 
These signals can be found in the nanoHertz GW frequency detectable by PTAs (\citealt{sbs19, PTAreview}). Current PTA sensitivity from the 11-year NANOGrav dataset has not yielded a detection of GWs from an individual SMBHB system \citep{NANOGrav11yr}, however, there is potential evidence for a GW background (GWB) in the NANOGrav 12.5-year data~\cite{GWB}. Such a GWB could arise naturally from the cosmic merging history of SMBHBs, and thus motivates searching for the signals of individual nearby binary systems. While major galaxy mergers which result in these SMBHBs are rare and as the GW-driven inspiral typically takes tens of millions of years: once such a system is detected, it will be in-band for millennia. Currently, there is compelling evidence that the brightest quasars are produced as a result of major galaxy mergers and represent the end product of the coalescence of central SMBHs from the individual merging sources \citep{Urrutia+2008,Treister+2010,Glikman+2015}.    

Surveys carried out by Square Kilometre Array (SKA) mid and low frequency instruments later this decade (along with ngVLA and DSA2000) are set to improve upon PTAs by discovering more millisecond pulsars \citep{keane15}; and improving the sensitivity by $\sim$ two orders of magnitude, corresponding to a minimum strain of $10^{-16}$ \citep{Xin}. This strain is detectable across the entire PTA frequency band, and subtraction of the GWB at low-frequencies is actively being improved upon. Here we refer to the SKA as both a tool for GW detection via its inclusion in PTA experiments, and also as an imaging tool on its own. Hereafter, we refer to the instrument's role in pulsar discovery and inclusion in PTAs as SKA-PTA. The Next Generation Very Large Array (ngVLA; \citealt{Reid2018}) and Deep Synoptic Array (DSA-2000; \citealt{DSA2000}) will also be valuable pulsar timing tools, as well as imaging tools for potential SMBHB host galaxies.

Here we evaluate the role of strong gravitational lensing in aiding and accelerating GW detection efforts and extracting multi-messenger information from SMBHBs. GWs are strongly lensed in a similar way to EM waves, enabling us to not only see a magnified image of the SMBHB host galaxy, but also detect an enhanced GW signal.
For example, assuming a conservative magnification factor of $\mu=3$, lensing improves the PTA detection volume by a factor of about $\sqrt{3^{3}}\sim 5$ for fixed GW strain and chirp mass of a given source. Additionally, for a fixed source redshift and GW strain, the $\sqrt{\mu}$ factor allows us to detect systems that are $\sqrt{3} $ times less massive in chirp mass, opening up the detectable parameter space for SMBHBs considerably.

Our goal is to determine how many SMBHBs could be strongly lensed within a certain co-moving volume and what additional multi-messenger information can be extracted from these systems. In particular, we explore dual quasars, or active galactic nuclei (AGN), since they may trace galaxy mergers and hence eventual SMBHB mergers \citep{Sanders88, Volonteri2003, Granato2004, Hopkins2008, Treister+2010, Goulding2019, Andrew}. 

In order to compute the number of strongly lensed SMBHBs in a redshift volume, we use the galaxy stellar mass function (GSMF; \citealt{Muzzin}) and quasar luminosity function (QLF; \citealt{QLF}, a binary fraction \citep{binfrac}, and a strong lensing probability (or optical depth; \citealt{Barnacka}). For both the QLF and GSMF we bin over $z$ and chirp mass, assuming a magnification factor $\mu$, and carry out a Monte Carlo exploration in GW frequency in order to compute the lensed strain, $h$. We thus determine which bins have detectable SMBHB systems for NANOGrav and the SKA. 

Intriguingly, if the SMBHBs are sufficiently massive and at high enough frequency, they may have an $\dot{f}$ which allows for the GW signal to evolve from one frequency bin to the next (of the order of 1-3 nHz). This would in turn allow us to observe SMBHBs motion in lensed images with e.g. very long baseline interferometry (VLBI) while simultaneously measuring the corresponding GW evolution.

Such strongly lensed SMBHBs would be a rich source of astrophysical information. Multiple images of an SMBHB with an expected time delay of years would allow us to directly observe the orbiting SMBHs for the first time, providing us with potentially millions of years of direct observations of the coalescing system. As we demonstrate in this work, it may even be possible to monitor the orbital dynamics of the binary, via its lensed images, to provide constraints on interactions with gas and stars that surround it. 

The outline of our paper is as follows: in Section~2, we briefly discuss the relevant strong lensing effects following up in Section~3 with a brief description of lensing of gravitational waves. In Section~4, we describe the methodology derived to calculate the number of expected lensed SMBHBs and their strains as well as their time delays. The detailed results for the detection prospects of these multi-messenger events are presented in Section~5, followed by conclusions and discussion.

\section{Strong Gravitational Lensing}

In the strong gravitational lensing regime, the deflection of light by a foreground mass distribution results in the production of multiple magnified images of a single background source. The lens' spatial distribution, alignment with respect to the source, and gravitational potential influence the caustics of the lensed system, resulting in varying degrees of magnification and distortion in shape, as well as image multiplicity (see \citealt{lensing_eqs,KneibPN2011,Barnacka} for lensing basics and details on the properties of lensed image positions and magnifications). The positions of the multiple images are determined by solving the lensing equation:
\begin{equation}
    \Vec{\beta} = \Vec{\theta} - \nabla \psi = \Vec{\theta} - \Vec{\alpha} \, ,
\end{equation}
where $\beta$ is the source position, $\theta$ denotes the image positions, $\psi$ is the 2-D lensing potential, and $\alpha$ is the scaled deflection angle of light waves or gravitational waves from the source \citep{Blandford+Narayan}.

The mass distribution of galaxy strong lenses is commonly modeled by either a singular isothermal sphere (SIS) or a singular isothermal elliptical (SIE) lens \citep{sigma_v}. The strength of a lens can be characterized by its Einstein radius, or the area of the lensing cross-section. For an SIS lens, the Einstein radius is defined by the point at which the lens's surface-mass density is equal to the critical density \citep{Blandford+Narayan, sigma_v}. If a source's position $\beta$ falls within the Einstein radius of an SIS, the lens produces a double-image of the source (and no additional de-magnified image at the core, since this model is not a \textit{cored} isothermal sphere) \citep{Blandford+Narayan}. 

For an SIE lens model, a quadrupolar component is added to either the lens potential or mass distribution, characterizing the caustics of the elliptical lens. Formally, the caustics are curves of infinite magnification in the image plane that indicate boundaries between regions of different image multiplicities \citep{Blandford+Narayan}. A background source that has crossed the outer caustic will exhibit two magnified images in the image (lens) plane. Furthermore, if this source crosses the inner caustic, it results in quadruple images \citep{Blandford+Narayan}. Sources located very close to caustics, or the corresponding critical lines in the source plane, experience the greatest magnifications \citep{Blandford+Narayan, Barnacka}.

\begin{figure}[b]
    \centering
    \includegraphics[width=0.8\textwidth]{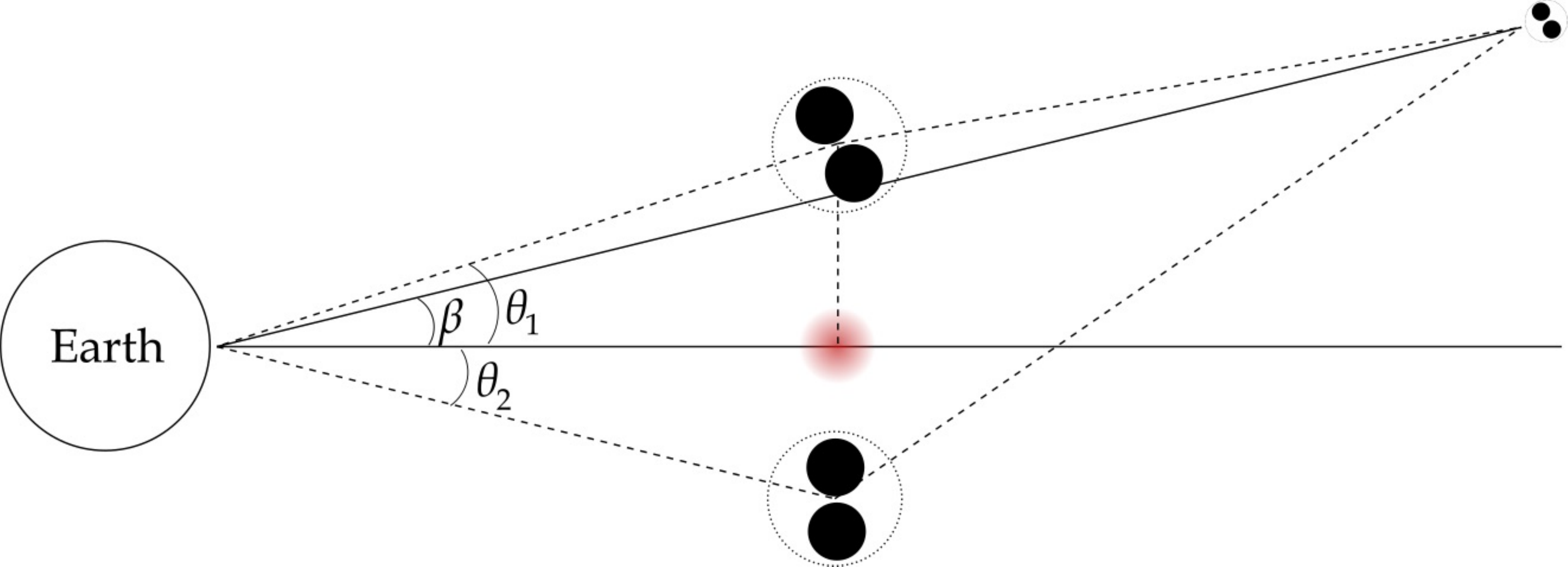}
    \caption{An single isothermal sphere (SIS) lensing model is a sufficient approximation of lensing configurations produced by galaxy scale lenses or even galaxy cluster scale lenses, as it simplifies the spatial distribution of the lens by introducing spherical symmetry \citep{Blandford+Narayan,sigma_v,Barnacka}. The lens is shown in red, $\beta$ is the position of the SMBHB in the source plane, $\theta_1$ \& $\theta_2$ are the image plane positions of the source images, between which there exists a time delay due to the difference in path length traversed by light. To produce two images as shown, the source must lie within the Einstein radius, $\theta_E$, of the lens. An SIS model can produce at most two magnified images of a background source.
    }
    \label{fig:lensconfig}
\end{figure}

Here, depicted in \autoref{fig:lensconfig}, we explore the possibilities of detecting strongly lensed SMBHBs in the SKA era using a simple SIS model for the lens, as in \cite{SIS, Barnacka}. In this configuration, at most two lensed images are produced with positions $\theta_1$ and $\theta_2$ as shown in \autoref{fig:lensconfig} (we ignore the de-magnified third image from cored models). The source position is given by $\beta$, and the magnification factor is equal to the ratio of the position of the images to that of the source: \(\mu_{1,2} = \frac{\theta_{1,2}}{\beta}\) \citep{SIS, lensing_eqs, Barnacka}.

\section{Strongly Lensed Gravitational Waves}

PTAs are poised to detect the low frequency GWB from the population of SMBHBs. Importantly, the presence of this GWB does not impede the detection of GW signals from individual continuous sources, since these are likely at much higher frequencies \citep{Xin}. The sky location-, inclination-, and polarization-averaged amplitude, or strain, of a lensed GW signal is given by \citep{hGW, Ezquiaga+2020}
\begin{equation}
    h = \frac{2\mathcal{M}_c^{5/3}(\pi f_{\text{GW}})^{2/3}}{d_L}\mu^{1/2} \, ,
    \label{eq:strain}
\end{equation}
where $\mu$ is the magnification factor, $\mathcal{M}_c$ is the chirp mass of the system, $f_{\text{GW}}$ is GW frequency, and $d_L$ is the luminosity distance to the source. Furthermore, the GW frequency evolves in time, denoted by $\dot{f}$:
\begin{equation}
    \dot{f} = \frac{96}{5}\pi^{8/3}\mathcal{M}_c^{5/3}f^{11/3} \, .
    \label{eq:fdot}
\end{equation}
This highlights the fact that the more massive the system and the greater its frequency, the more rapidly the GW signal evolves.

\section{Methodology to estimate the number of lensed SMBHBs}

\begin{figure}[h!]
    \centering
    \includegraphics[width=\linewidth]{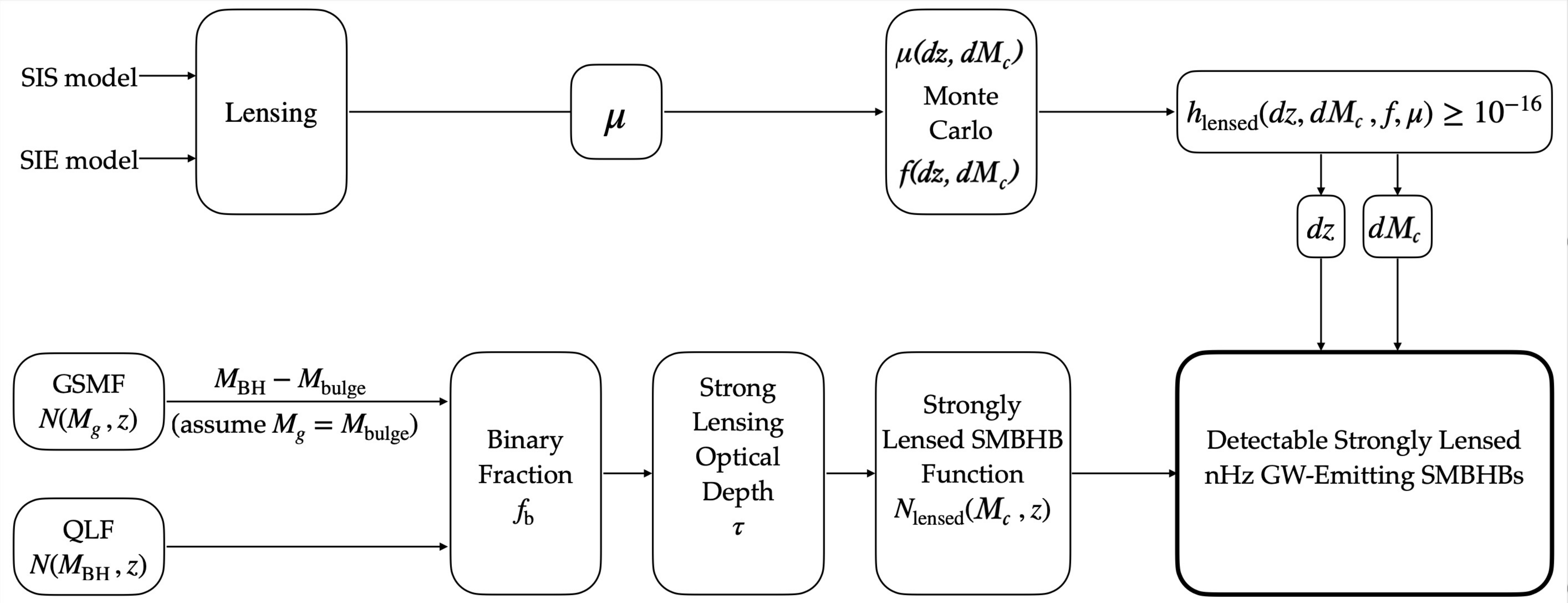}
    \caption{Framework for our calculations. We first determine the parameter space for which lensed SMBHBs produce detectable GWs by performing a Monte Carlo in frequency and magnification space in 2D redshift and chirp mass bins. We populate these bins with SMBHBs using mass functions constructed from a GSMF and QLF with an additional redshift-dependent binary fraction and the strong lensing probability. We integrate over all detectable bins to find the cumulative number of detectable, strongly lensed SMBHBs within $z=2$.}
    \label{fig:methods}
\end{figure}

We develop the methodological framework, presented in \autoref{fig:methods}, to calculate the distribution of detectable, strongly lensed SMBHBs within a specified redshift volume. We estimate which $z$---$\mathcal{M}_c$ bins host SMBHBs that will emit detectable GWs and populate them using a mass function for individual SMBHs, applying binary fraction and strong lensing probability cuts. We integrate over this detectable parameter space to compute the cumulative redshift distribution of detectable, strongly lensed GW-emitting SMBHBs.

\subsection{Strong Lensing Probability}
We assume an SIS lens model for foreground galaxy lenses that will magnify our background source population of SMBHBs. This simple mass model has been shown to offer a good fit to lensing galaxies and the simplicity of the profiles permits easy calculation of the strong lensing probability  \citep{SISapprox, Barnacka}. The differential strong lensing probability is given by
\begin{equation}
    d\tau = \frac{F(1+z_L)^2(\frac{D_{L}D_{LS}}{D_{H}D_{S}})^2}{\sqrt{\Omega_M(1+z_L)^3 + (1-\Omega_M-\Omega_{\Lambda})(1+z_L)^2 + \Omega_{\Lambda}}}dz_L \, ,
    \label{eq:dtau}
\end{equation} 
where $D_L, D_S, \text{ and } D_{LS}$ are the angular diameter distances to the lens, to the source, and between the lens and source, respectively; $D_H$ is the Hubble distance; $z_L$ is the redshift position of the lens; $\Omega_M \text{ and } \Omega_{\Lambda}$ are the mean matter density and cosmological constant, respectively; and F measures the effectiveness of creating double images, chosen by \cite{Fukugita1992} to be 0.047 \citep{Zeldovich1964, DryerRoeder1973, TurnerGott1984, Fukugita1992, Barnacka}. As per the definition, we integrate over all possible positions between observer and source to compute the number of lenses that fall within the Einstein angle cross-section along the line of sight to the source. For this reason, $\tau(z_s) = \int_0^{z_s} d\tau$ is called the strong lensing optical depth, synonymous with strong lensing probability \citep{Barnacka}.

The distributions of lenses are incorporated into the $F$ parameter in \autoref{eq:dtau}, which reports the effectiveness of producing double images \citep{Fukugita1992, Barnacka}. We adopt the value for $F$ from \citealt{Fukugita1992}, whose authors assume a lens population that follows quasar catalogs (they specifically use \cite{HB} and \cite{Boyle+1990} as examples of their predictions) and has SIS-approximated mass distributions.

\subsection{Binary Fraction}

The binary fraction expresses the likelihood of finding a binary system of SMBHs at the center of a galaxy. \cite{binfrac} present a pair fraction evolving with redshift: \(f_p(z) = 0.07(1+z)^{0.6}\). We convert this power law into a binary fraction
\begin{equation}
    f_b(z) = 0.02(1+z)^{0.6}
    \label{eq:bin_frac}
\end{equation}
by changing the pre-factor from 0.07 to 0.02 --- motivated by predictions of a 1-4\% probability for massive galaxies hosting SMBHBs \citep{BS2011, 0.02} --- assuming that the binary fraction evolves over redshift in a similar manner to that of the pair fraction.

\subsection{Mass Functions}

We compare results from calculations using two different mass functions for SMBHs. One is starting from host galaxy properties via the galaxy stellar mass function (GSMF; \citealt{Muzzin}) built from a catalog of both star-forming and quiescent galaxies of various morphologies. 
The other is from the SMBH mass function derived from a quasar luminosity function (QLF; \citealt{QLF}).

\subsubsection{Black Hole-Galaxy Scaling}

Since the strain $h$ is a function of chirp mass, we need to be able to convert between $\mathcal{M}_c$ and the mass function inputs, e.g. total black hole mass, $M_{\text{BH}}$, and galaxy stellar mass, $M_g$. We present this work as a proof of concept calculation, so we assume the optimal case mass ratio of $q=1$ such that $\mathcal{M}_c^{5/3} = q/(1+q)^2 M_\mathrm{BH}^{5/3} = \frac{1}{4}M_\mathrm{BH}^{5/3}$ and defer a more complex analysis in $q$ to future work. $M_\mathrm{BH}$ is the sum of two equal mass BHs. To compute $M_g$ from $M_\mathrm{BH}$, we use the $M_{\text{BH}}$---$M_{\text{bulge}}$ scaling relation modeled in \cite{MMbulge}, for which we assume that $M_g = M_{\text{bulge}}$.

\subsubsection{Deriving source counts from the Galaxy Stellar Mass Function}

\cite{Muzzin} construct a GSMF from a sample of 95,675 galaxies in the COSMOS/UltraVISTA survey. We use the total mass function arising from the sum of independent fits for quiescent and star-forming galaxies. The total GSMF, $\phi(M_g)$, is defined between $z = 0.2$ and $z = 4$, and a Schechter fit is produced for each of the 7 redshift bins. A double-Schechter function form is used to fit the mass function at low redshift bins $z<1.0$. We interpolate along the redshift bins as per the prescription laid out in \cite{Andrew} to produce a continuous $\phi(M_g, z)$. To obtain a number density of galaxy sources from this GSMF, we integrate over redshift and galaxy mass:
\begin{equation}
    n(M_g, z) = \int_{z}^{z+\,dz} \int_{M_g}^{M_g+\,dM_g} \phi(M_{g}, z) \,dM_g \,dz \, ,
    \label{eq:ndens_GMF}
\end{equation}
where $n$ is the number density of galaxies with stellar masses between $M_g$ and $M_g+\,dM_g$ found within $z$ and $z+\,dz$. The total number count of such galaxies is
\begin{equation}
    N(M_g, z) = n(M_g, z)V_c\, ,
    \label{eq:N_GMF}
\end{equation}
where $V_c$ is the co-moving volume between $z$ and $z+\,dz$. 

\subsubsection{Deriving source counts from the Quasar Luminosity Function}

A complementary approach to using a GSMF for the host galaxies and deploying scaling relations for the central SMBHs is to directly use a measured QLF \citep{QLF}. The QLF, $\phi(L)$, is a double power law fit to the observational data compiled in \cite{QLF}, with the form
\begin{equation}
    \phi(L) = \frac{\phi_*}{(L/L_*)^{\gamma_1} + (L/L_*)^{\gamma_2}}\, ,
    \label{eq:QLF}
\end{equation}
with normalization $\phi_*$, break luminosity $L_*$, and faint and bright-end slopes $\gamma_1$ and $\gamma_2$, respectively. The authors construct the same observed bolometric QLF from a merger- and BH-growth-driven quasar light curve model, where the QLF is built from a convolution of the quasar formation rate and differential lifetime.
Extracting the formation rate and multiplying by a factor of $dt/dz$ to write a mass function $\phi(M_{\text{BH}}, z)$, we can calculate the number density of BHs similarly to \autoref{eq:ndens_GMF}:
\begin{equation}
    n(M_{\text{BH}}, z) = \int_{z}^{z+\,dz} \int_{M_{\text{BH}}}^{M_{\text{BH}}+\,dM_{\text{BH}}} \phi(M_{\text{BH}}, z) \,dM_{\text{BH}} \,dz \, ,
    \label{ndens_QLF}
\end{equation}
where $n$ is the number density of quasars with BH masses between $M_{\text{BH}}$ and $M_{\text{BH}}+\,dM_{\text{BH}}$ found within $z$ and $z+\,dz$.
The total number count of such sources of quasars is 
\begin{equation}
    N(M_{\text{BH}}, z) = n(M_{\text{BH}}, z)V_c\, .
    \label{eq:N_QLF}
\end{equation}
We describe the construction of the black hole mass function (BHMF) in greater detail in the Appendix.

\subsection{Detectable Source Population Estimates} 
\label{sec:MCs}

We bin over $z$ and $\mathcal{M}_c$ to determine which bins will count towards our final population of detectable, strongly lensed SMBHBs. We perform Monte Carlo (MC) calculations to determine which $z\mathrm{-}\mathcal{M}_c$ bins give rise to SMBHBs with $h \geq 10^{-16}$ and populate them using \autoref{eq:ndens} and \autoref{eq:N_lensed}. Summing over all detectable bins determined by the MC, we produce a cumulative distribution of such strongly lensed detectable systems.

\subsubsection{Frequency Sampling}

Within each realization of our MC for every $i^\mathrm{th}$ bin, we sample over time to coalescence with
\begin{equation}
    t_c(\mathcal{M}_c, f) = \frac{5}{256}(\pi f)^{-8/3}\mathcal{M}_c^{-5/3} \, ,
    \label{TTC}
\end{equation}
from \cite{FandTTC}, and assign the binaries in the bin a value $t_{c, i}$ from a uniform distribution between 0 years and $t_c$ from 1 nHz for a given chirp mass. From this $t_{c,i}$ we compute the GW frequency, $f_i$. 

As expected, most realizations produce a low-frequency binary, resulting in a highly skewed distribution peaking around $f = 1$ nHz. The greater the GW frequency of the binary, the quicker it evolves (as per \autoref{eq:fdot}) and the closer it is to coalescing, so binaries spend most of their lifetimes at low frequency. 

\subsubsection{Magnification Sampling} \label{sec:mu_sampling}
Within each realization of our MC for every $i^{\text{th}}$ bin, we also sample over a $\mu$ probability density function (PDF) in order to use a realistic magnification factor in our $h_i$ calculations. We adopt a PDF from \cite{Dai+2017}, taking the form of:
\begin{equation}
    p(\mu) = A(t_0)\int_0^{+\infty}\left( {e^{\frac{\lambda}{t+t_0}}-2t}\right) \times\frac{1}{\sqrt{2\pi}\sigma}e^{-\frac{(\ln{\mu}-\delta-t)^2}{2\sigma^2}} \,dt \,,
    \label{eq:mu_pdf}
\end{equation}
where $A(t_0)$ is the normalization, and $t_0$, $\sigma$, and $\delta$ are fit parameters that regulate the distribution tail, the log-normal distribution width, and the shift, respectively. The authors choose $\lambda=5$, which provided the most realistic fit. This probability distribution varies with redshift, as we expect more distant sources to have a greater likelihood of being strongly lensed and by greater factors. We interpolate between the fixed-redshift best fit parameters that the authors report in order to compute the smoothly $z$-varying PDF $p(\mu;z)$ from \autoref{eq:mu_pdf}:
\begin{equation}
    p(\mu;z) = A(t_0(z))\int_0^{+\infty}\left( {e^{\frac{\lambda}{t+t_0(z)}}-2t} \right) \times \frac{1}{\sqrt{2\pi}\sigma(z)}e^{-\frac{(\ln{\mu}-\delta(z)-t)^2}{2[\sigma(z)]^2}} \,dt \,.
    \label{eq:mu_pdf_z}
\end{equation}
We sample from $p_i(\mu;z_i)$ in each realization of each bin to obtain a physical $\mu_i$ for the lensed strain calculation.

\subsubsection{Measuring Detectability}
From $f_i$ and $\mu_i$, we compute the lensed strain, $h_i$, with \autoref{eq:strain} using the redshift and chirp mass values of the particular $i^{\text{th}}$ bin. We evaluate the peak value of the $h_i$ distribution. 
If this $h_i$ $\geq 10^{-16}$, we determine that the lensed SMBHBs found within the $i^{\text{th}}$ bin are detectable, so we populate the bin. Otherwise, the bin's binaries are excluded from the count. Summing over all detectable bins determined by the MC, we produce a cumulative distribution of such strongly lensed detectable systems.

\subsection{Lensed Binary Number Density}

To obtain a census of the population of detectable strongly lensed SMBHBs, we integrate over all determined detectable $z\mathrm{-}\mathcal{M}_c$ space the product of the mass function from \cite{Muzzin} or \cite{QLF}, with the evolving binary fraction from \autoref{eq:bin_frac}, and strong lensing probability from \autoref{eq:dtau} as shown below:
\begin{equation}
    n_{\text{lensed}} = \int_{0}^{z} \int_{M_{\text{min}}(z')}^{M_{\text{max}}(z')} \phi(M, z') f_b(z') \tau(z') \,dM \,dz'\, ,
    \label{eq:ndens}
\end{equation}
where $\phi$ is our chosen mass function; $f_b$ is our evolving binary fraction; $\tau$ is the lensing probability; and $M_{\text{min}}$ and $M_{\text{max}}$ are interpolated functions of redshift describing, respectively, the minimum and maximum masses of sources producing detectable lensed strains. In a similar manner as in \autoref{eq:N_GMF} and \autoref{eq:N_QLF}, the final number count of detectable, strongly lensed SMBHBs is given by
\begin{equation}
    N_{\text{lensed}} = n_{\text{lensed}}V_c\, .
    \label{eq:N_lensed}
\end{equation}
Importantly, we compute the \textit{detectable} number of lensed events, as we count galaxies/quasars only in the detectable lensed parameter space determined by the frequency and magnification MC strain calculations outlined in \autoref{sec:MCs}.

\subsection{Time Delays}

A critical aspect of the strong lensing regime is the production of multiple images of the source. 
The different geodesic paths taken by GWs (and light) to arrive to Earth introduce time delays between signals from each image.

In the case of SIS lens models, as noted previously up to two images of the source can be produced. \cite{Barnacka} provides the calculation for the Einstein radius of a SIS that we adopt:
\begin{equation}
    \theta_{E} = 4\pi\frac{\sigma_v^2}{c^2}\frac{D_{LS}}{D_{OS}} = 1.15\left(\frac{\sigma_v}{200 \text{ km/s}}\right)^2\, ,
    \label{eq:eradius}
\end{equation}
where $D_{LS}$ is the angular diameter distance between the lens and the source, $D_{OS}$ is the angular diameter distance between the observer and the source, and $\sigma_v$ is the velocity dispersion of the lens. To produce double images, the source must lie within the Einstein radius of the SIS lens along its line of sight. Thus, as long as $\beta < \theta_E$, the positions of the images are given by \citep{lensing_eqs, Barnacka}
\begin{equation}
    \theta_{A,B} = \beta \pm \theta_E\, .
    \label{eq:imagepos}
\end{equation}
We can use these positions to calculate the time delay, $\Delta t$, between the images:
\begin{equation}
    \frac{2c\Delta t}{(1+z_L)} = \frac{D_{OS}D_{OL}}{D_{LS}}(\theta_B^2 - \theta_A^2)\, ,
    \label{eq:timedelay}
\end{equation}
where $z_L$ is the lens redshift and $D_{OL}$ is the angular diameter distance from the observer to the lens \citep{timeDelay, Barnacka}.

\section{Results}
\subsection{Detectable Parameter Space}

\begin{figure*}[h!]
\centering
\subfloat{\includegraphics[width=0.3135\textwidth,valign=c]{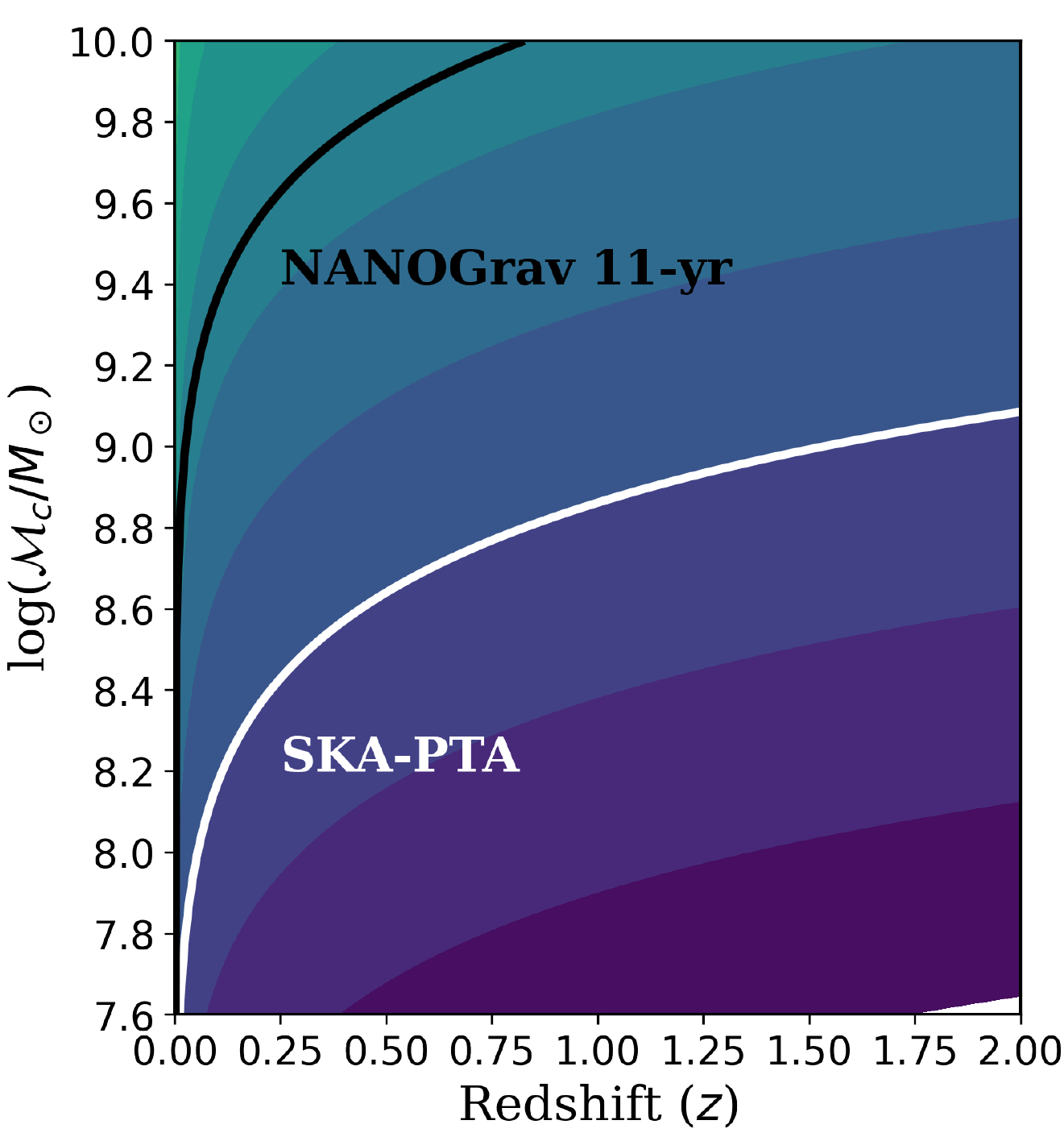}}
\hfill
\subfloat{\includegraphics[width=0.275\textwidth,valign=c]{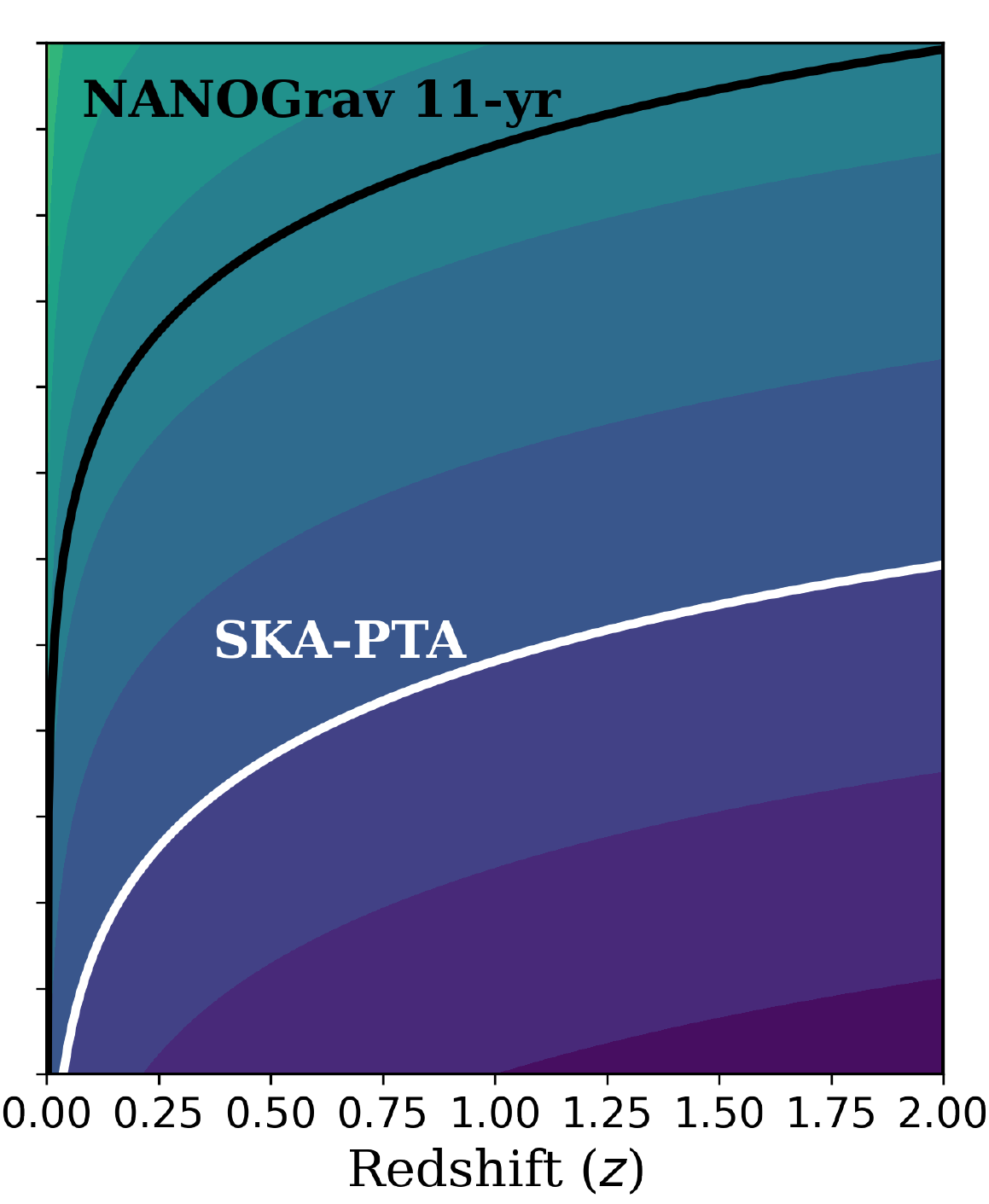}}
\hfill
\subfloat{\includegraphics[width=0.275\textwidth,valign=c]{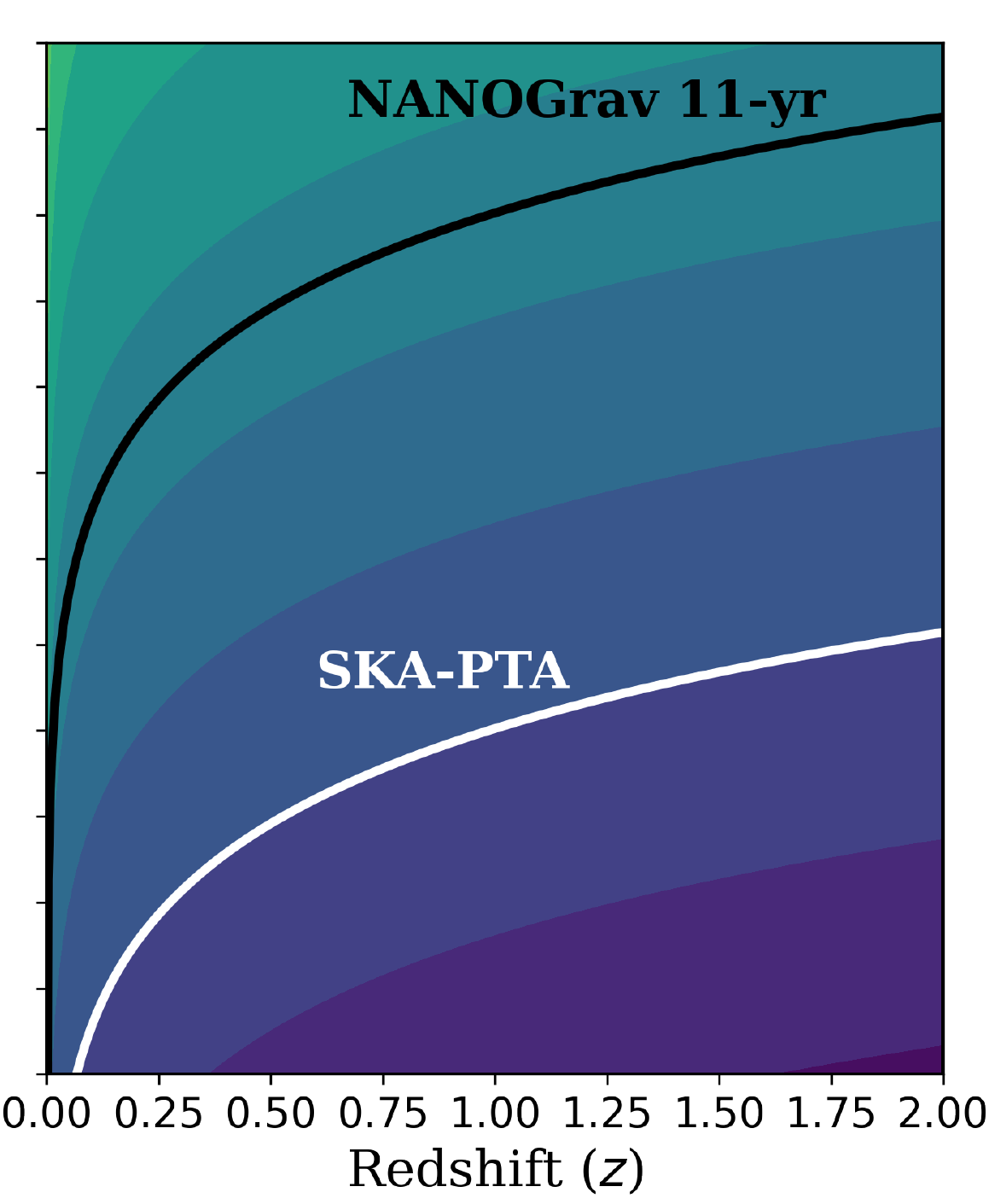}}
\hfill
\subfloat{\includegraphics[margin=0cm 0cm 0cm -1.25cm, width=0.0705\textwidth,valign=c]{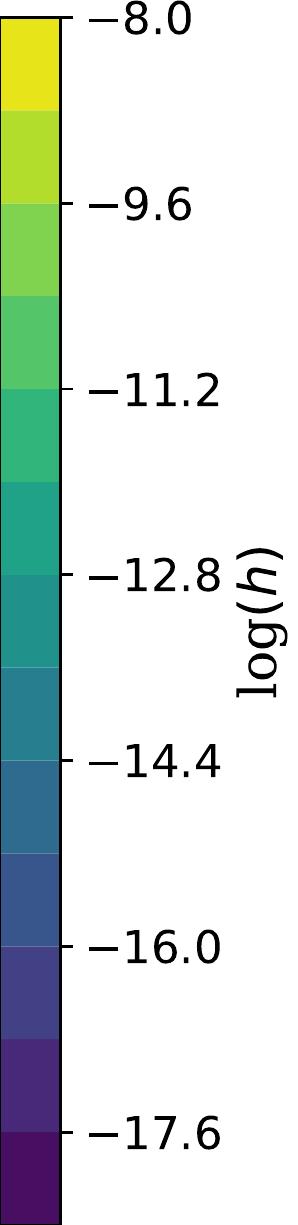}}

\caption{Supermassive black hole binary systems may be strongly lensed, and are detectable by current and future PTAs, e.g. NANOGrav and the SKA-PTA. Here we show three example strong lensing detection scenarios, resulting from targeted searches for SMBHBs with $f_\mathrm{GW} = 13$ nHz. We perform a Monte Carlo over $\mu$ and $f_\mathrm{GW}$ in our later analysis. The solid black curve is the current strain upper limit from NANOGrav~\citep{NANOGrav11yr}, while the white contour marks the future strain sensitivity with SKA~\citep{Xin}. \emph{Left Panel:} SIS lens model with modest magnification of $\mu=3$. Pessimistic interpretation of parameter space for detectable GWs from strongly lensed SMBHBs. GWs lensed by a SIS, result in the production of double images and comparatively low magnification. \emph{Central Panel:} SIE lens model with high magnification of $\mu=30$. Greater magnification and image multiplicities are expected from the average SIE lens model. \emph{Right Panel:} SIE lens model with rare exceptional magnification of $\mu=100$. Optimistic detectable parameter space for strongly lensed GW signals. Best-case magnification and greater image multiplicity due to caustic configuration from SIE model.}
\label{fig:param_map}
\end{figure*}

Magnification of the strain $h$ due to strong lensing broadens the parameter space for PTA-detectable SMBHBs.
As shown in \autoref{fig:param_map}, we present three example cases to demonstrate detection improvements from strong lensing. For each case, we use the SIS and SIE lens models and compute the corresponding $\mu$'s. For multi-messenger GW searches, knowledge of the sky location of a GW source (e.g. 3C 66B; \citealt{3c66b}) improves PTA sensitivity by a factor of 2 for targeted searches. We fold this improvement into our \autoref{fig:param_map} maps. We choose $f_{\text{GW}} = 13$ nHz to reflect the approximate GW frequency of periodic light curve SMBHB candidates in the Catalina Real Time Transient
Survey (CRTS) \citep{Xin}. 

The characteristic parameters for the first case are an SIS lens model with a modest magnification of $\mu = 3$. This is about the mean of the $\mu$ distribution for observed massive cluster lenses \citep{mu3}. The second case is an SIE lens, or a cluster lens (modeled by an SIE profile) with superb alignment geometry for the source and lens, resulting in significantly higher magnification values of $\mu = 30$ \citep{Bergamini+2022}. The third case is an SIE lens model with the rare chance alignment with $\mu = 100$. 
In fact, while we claim an optimistic magnification $\mu=100$ results from fortuitous lens-source alignment, a strongly lensed AGN has been detected at $z = 3.273$ with an estimated magnification factor of $\mu = 300$ \citep{Spingola+2019, mu300}. Therefore, though rare, such systems with magnifications on the order of hundreds do exist, and they will be the most crucial and exciting systems for us to explore. Furthermore, for this system, the large offset amplification from the high-magnification caustic configuration of the source and lens allows the authors to reconstruct a binary source of emission from the image configuration \citep{Spingola+2019, mu300}. As we argue below, even the detection of a  single such system as a multi-messenger event stands to vastly improve our current understanding of SMBHBs and physics of their mergers.

In this work, we focus primarily on the lensing signals produced by the SIS model. However, it is important to note that the SIE model results in caustic lensing configurations that produce the greatest magnifications with $\mu \sim 100$. We defer the more nuanced and detailed geometry of the SIE calculations to future work.

\subsection{The Strongly Lensed SMBHB Population}
\begin{figure*}[t]
    \centering
    \includegraphics[width=0.4\linewidth]{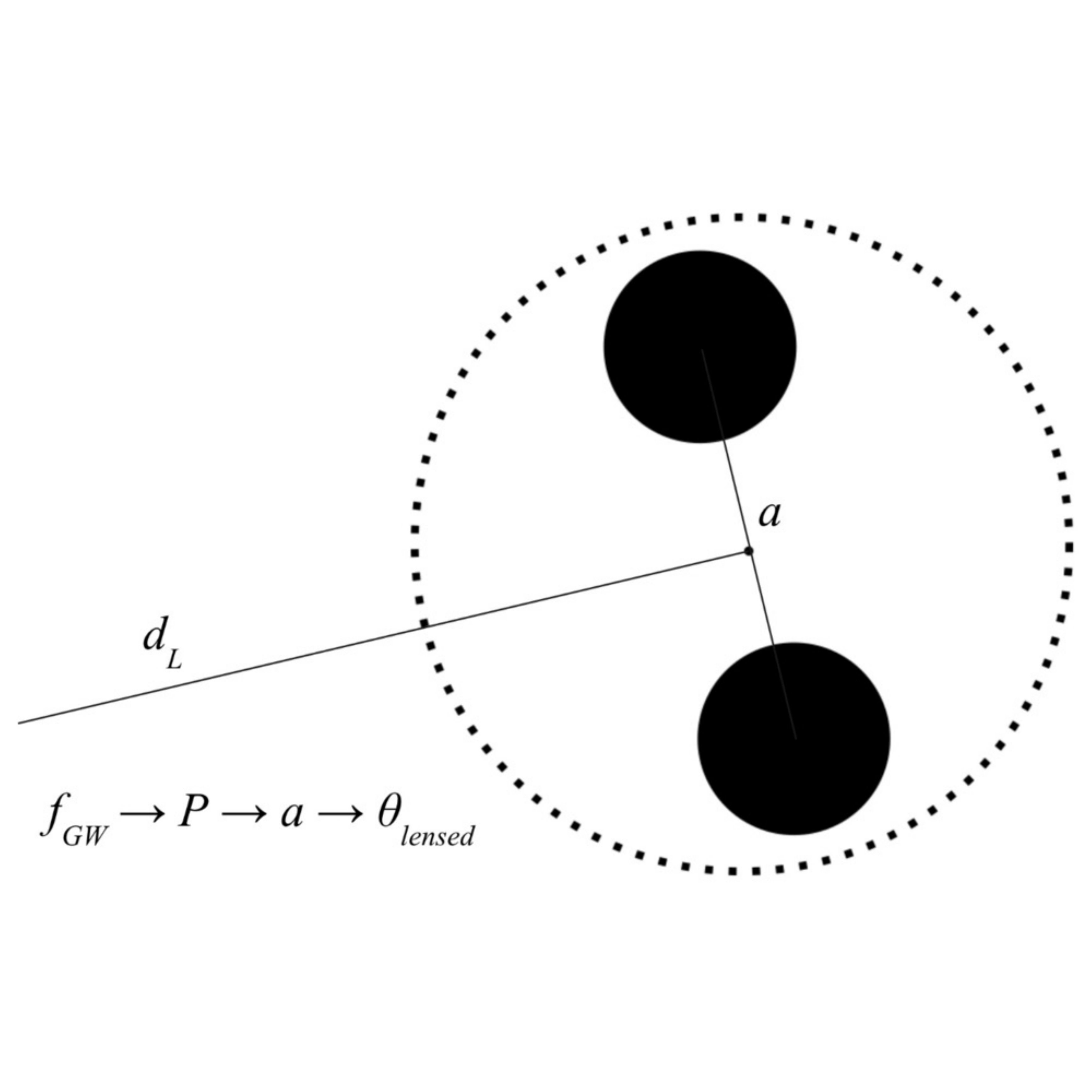} 
    \includegraphics[width=0.59\textwidth]{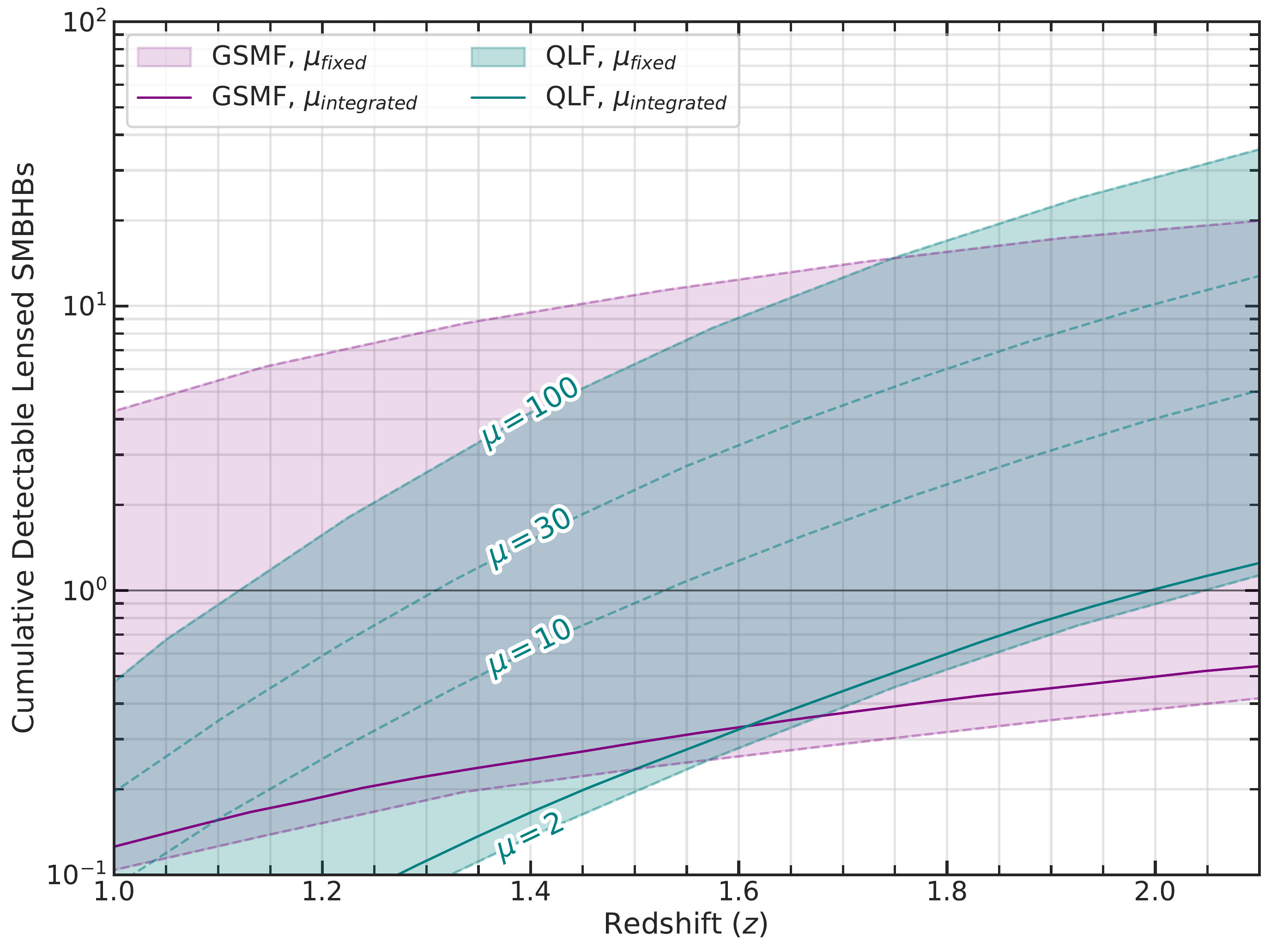}
    
    \caption{Left: A detailed view of a SMBHB lensed image. Using the magnification factor $\mu$ and the corresponding offset amplification factor \citep{Barnacka}, GW frequency, distance, and mass of the binary, we compute $a$, the semi-major axis using Kepler's third law. From $a$ we compute the angular separation, $\theta_{\text{lensed}}$, from which we infer the resolution capabilities of different instruments. Right Panel: Cumulative population of detectable lensed SMBHBs computed using both the QLF and GSMF. The wide bands are collections of fixed-magnification curves such that all sources are magnified by the same factor $2 \leq \mu \leq 100$ (dashed lines give examples). The solid curves are produced from integrating over the wide bands by sampling from a $\mu$ PDF \citep{Dai+2017} to obtain the appropriate $\mu_i$ for each $z\mathrm{-}\mathcal{M}_c$ bin, see Section \ref{sec:mu_sampling}. The positions of the integrated curves with respect to the bands reflect the greater likelihood of lower magnifications. While higher fixed $\mu$ yields a greater number of lensed events, the expected distribution based on a realistic $\mu$ distribution still results in $\sim$ few detectable lensed events within $z=2$ using the QLF calculation.
    }
    \label{fig:Nlensed_mu}
    
\end{figure*}

We use the SIS lens as a good approximation to model the mass distribution of the putative lenses, even though it yields smaller factors of $\mu$ \citep{Barnacka}. \autoref{fig:Nlensed_mu} shows the cumulative number of detectable, strongly lensed SMBHBs with MC-assigned GW frequencies and $\mu$ out to $z=2$. We plot two types of calculations: 1) We assume lensed sources will be magnified by a fixed $\mu$ and consider a range from 2 to 100, and 2) We count detectable sources based on a physically sampled $\mu$ as described in Section \ref{sec:mu_sampling}. The wide bands and solid curves represent the fixed $\mu$ range and the integrated $\mu$, respectively.

Interestingly, our calculations show that we can expect to find a few such sources within $z=2$ (and optimistically, with fortuitous alignment, within $z \approx 1.2$) using the QLF for both the physical scenario and the majority of the fixed $\mu$ range.
Using the GSMF, our integrated $\mu$ lensed SMBHB population does not give rise to a detectable source within $z=2$, although caustic configurations yielding greater $\mu$ could result in at least one detectable SMBHB. The difference in the steepness of the two distributions mirrors the shapes of the mass function models used and is also reflected in the top panel of Fig. 3 in \cite{Andrew}.
The distribution produced with the GSMF presents a lower limit on our detectable source population. However, we prefer the results using the QLF  because we believe quasars to be a more robust signpost for SMBHBs \citep{Sanders88, Volonteri2003, Granato2004, Hopkins2008, Treister+2010, Andrew}.

For a volume of $z \leq 2$, we count $\sim$800 detectable \textit{unlensed} binaries using the QLF. From the integrated $\mu$ result in \autoref{fig:Nlensed_mu}, we can expect to detect an additional $\sim$ few SMBHBs due to lensing. It is important to note that the first such source to be detected will likely have much greater $\mu$ due to magnification bias. While the proposed lensing scenario seems rare given our conservative assumptions and calculations, it is realistic nonetheless. Consider for example a SMBHB with $\mathcal{M}_c = 10^9 M_\odot$ and $f_{\text{GW}}=1$ nHz (c.f. \autoref{eq:fdot}, binaries spend the majority of their lifetimes at low frequencies). Unlensed, this binary would be detectable to $z \approx 0.25$, alongside some of the 800 unlensed binaries detected by $z \approx 0.2$. With $\mu=2$ it is detectable at $z \approx 0.3$, and with a strong magnification of $\mu=100$, its GWs would be detectable to $z \approx 1.5$. It is therefore possible that a strongly lensed binary would be amongst the first detected by PTAs.

In an effort to validate our results, we compute the estimated contribution to the 
GWB signal from our predicted underlying unlensed and lensed SMBHB population. A signal contribution from a few additional strongly lensed SMBHBs, as per the integrated $\mu$ result, will not realistically affect the total GWB characteristic strain unless they are incredibly loud. The maximum lensed contribution to the GWB based on our results is the combined signals from the $\sim$ 30 additional sources in the case of fixed $\mu=100$. We compute the characteristic strain of these contribution via $h_c^2 = \sum_i{h_i^2f_iT}$ \citep{hc}, where $h_i$ is the strain of the $i^{\text{th}}$ binary, $f_i$ is its frequency, and $T$ is the pulsar observation time (we use $\sim$16 years). We calculate $h_c$ due to our population of unlensed SMBHBs to be $\sim 2.05 \times 10^{-15}$, and we find the $h_c$ of the lensed contribution to be $\sim 3.88 \times 10^{-16}$, which is small compared to the predicted NANOGrav 12.5-yr range for total $h_c$ of $1.37$$-$$2.67 \times 10^{-15}$ (5\%$-$95\%; \citealt{GWB}).
We can use this GWB amplitude calculation to help constrain the binary fraction pre-factor in \autoref{eq:bin_frac}.

We present our results --- an expected detection of at least 1 additional detectable SMBHB due to an SIS strong lensing model --- as a proof of concept calculation. However, while the integrated curve gives rise to fairly small $\mu$ factors, it is important to note that the first lensed events we detect will likely have $\mu$ on the order of tens, maybe hundreds, due to magnification bias in observation and a greater probability of detection due to a caustic configuration \citep{Barnacka}. In the right panel of Fig. 17, \cite{Barnacka} shows that, for lensed pair events in caustic configurations, the sources experience offset amplifications that depend on their total magnification and the pairs' perpendicularity to the caustic. These sources will be most compelling, as they will exhibit the greatest flux magnifications and offset amplifications. 

We investigate the wealth of the multi-messenger astrophysics insights we can extract from the detection of single strongly lensed, higher-magnification SMBHB systems. In particular, we explore the idea of a ``golden'' binary that would be resolvable both electromagnetically and gravitationally. Such a binary would have resolvable EM counterparts within each lensed image such that we could even witness the orbit of the discrete SMBHs shrink. There could also be a significant enough time delay between the images such that PTAs could potentially track the GW signal evolution between images.

\subsection{Resolvability of EM Counterparts}
\begin{figure*}[b]
\centering
\subfloat{\includegraphics[width=0.3135\textwidth,valign=c]{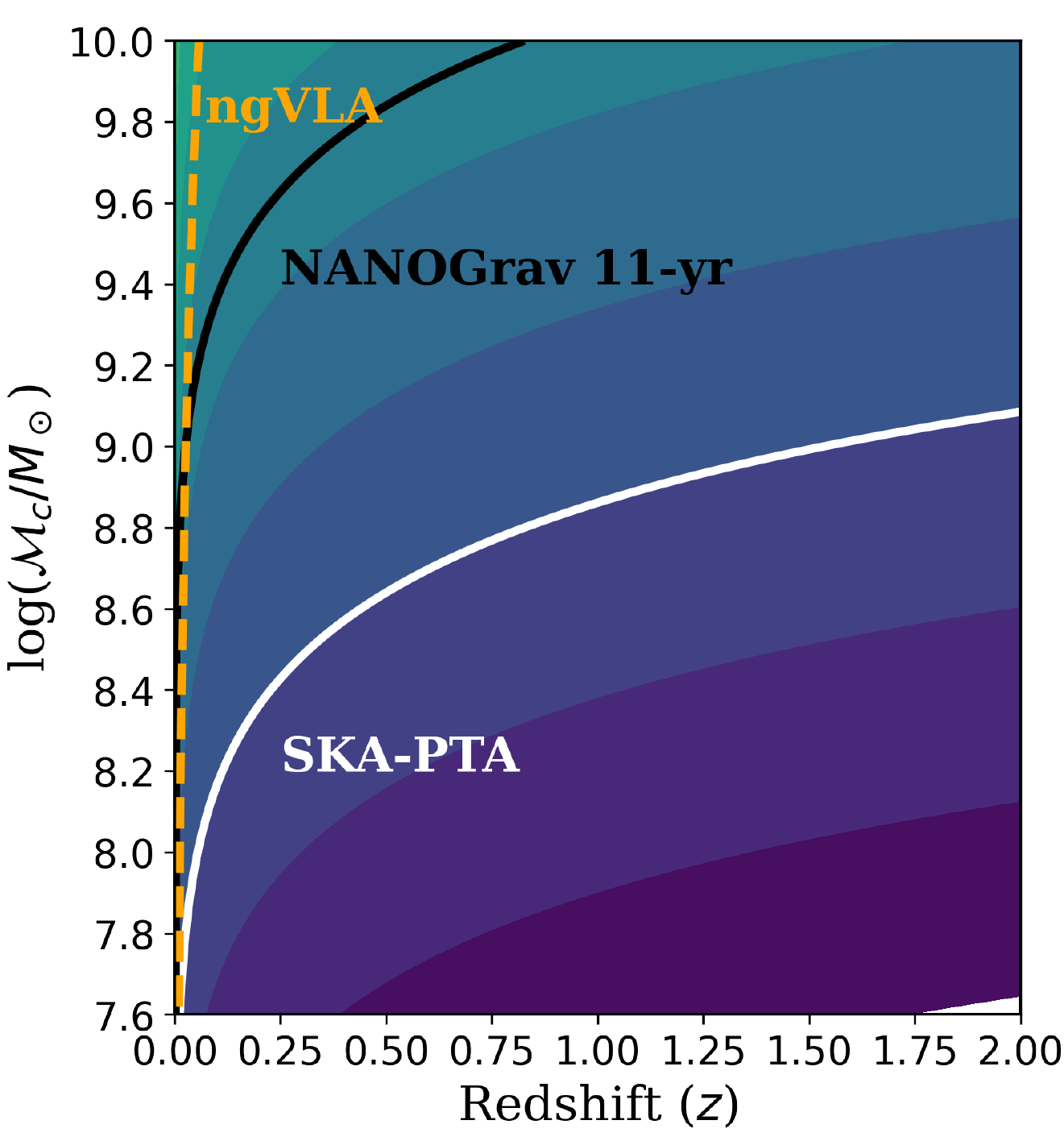}}
\hfill
\subfloat{\includegraphics[width=0.275\textwidth,valign=c]{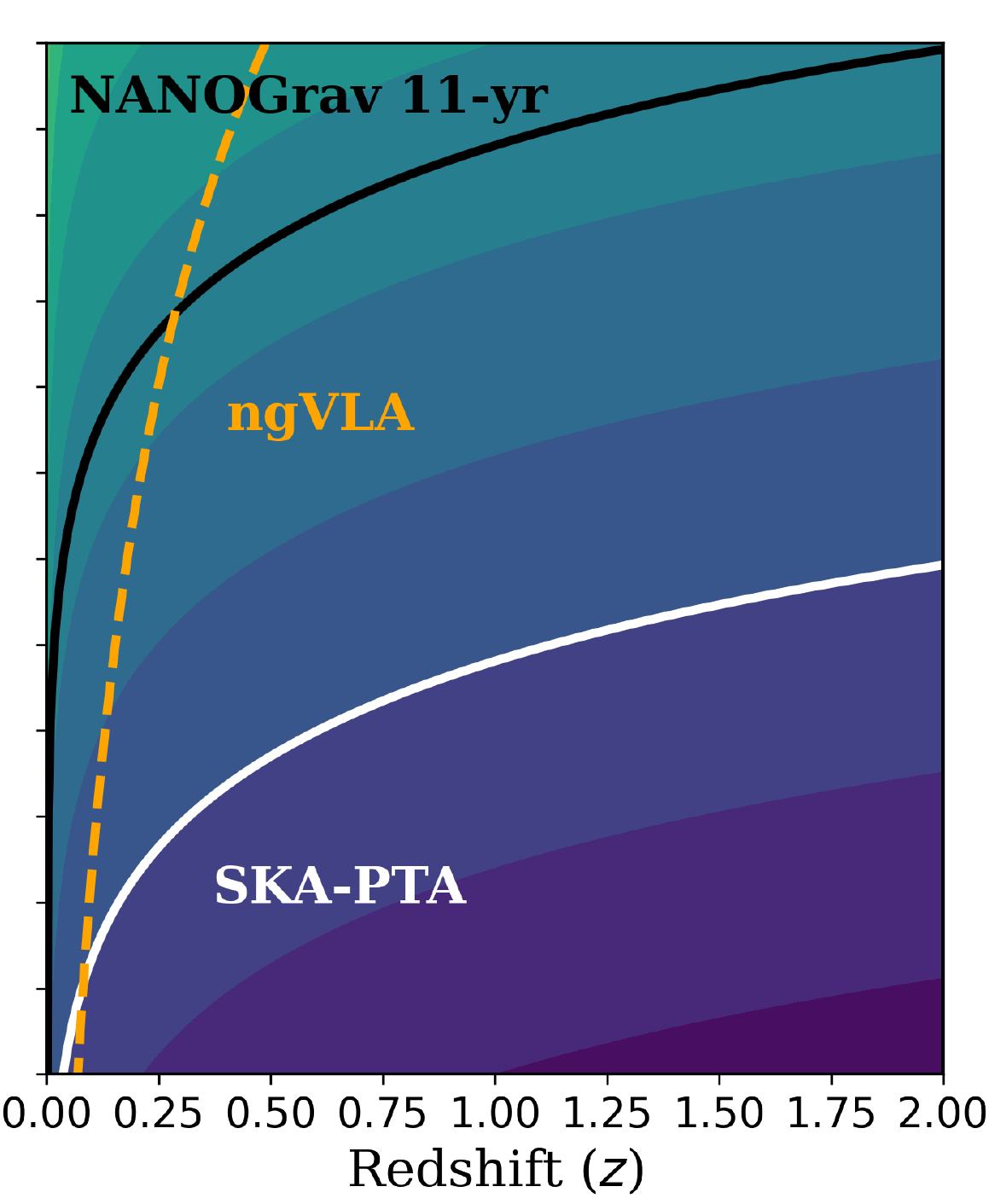}}
\hfill
\subfloat{\includegraphics[width=0.275\textwidth,valign=c]{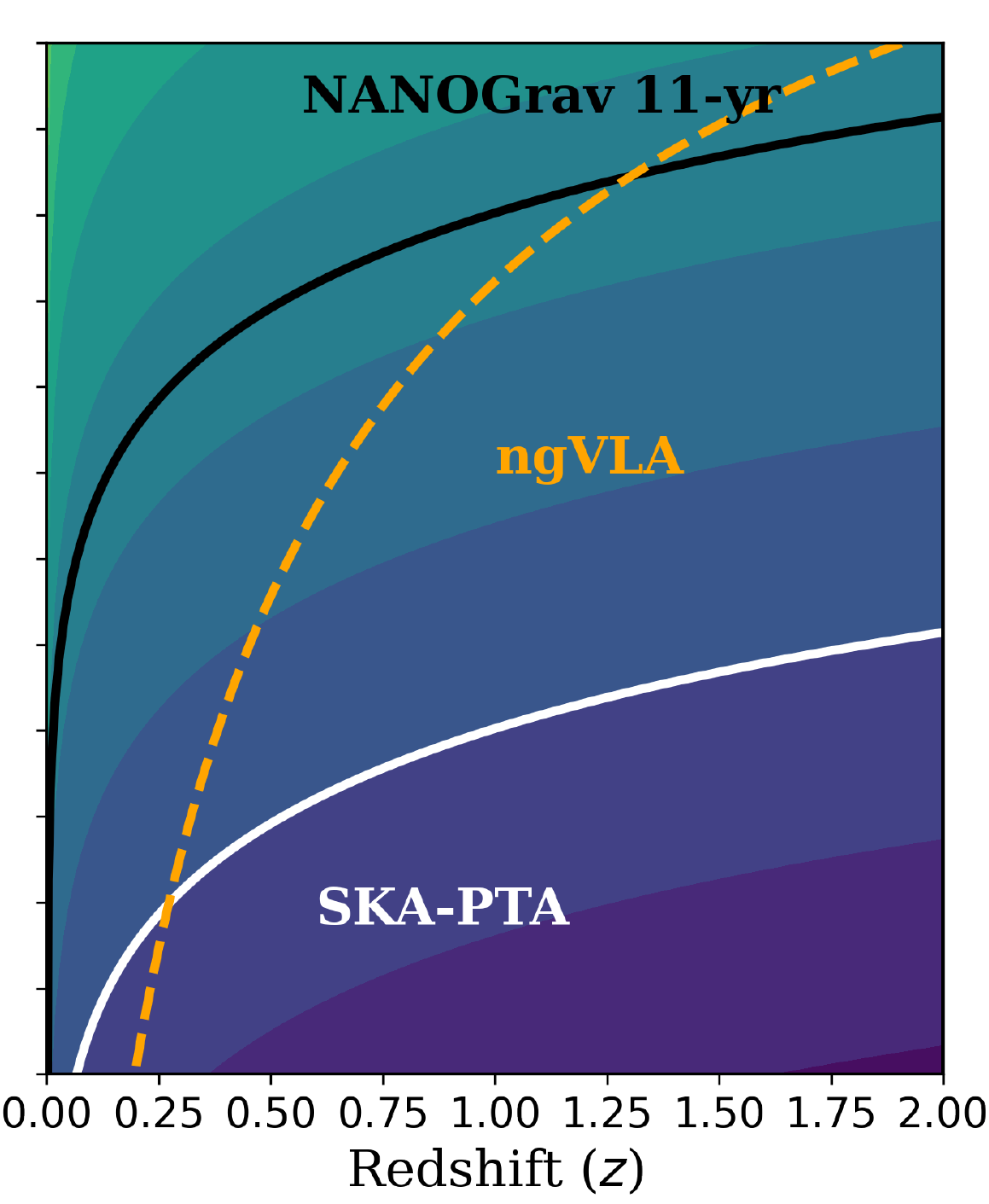}}
\hfill
\subfloat{\includegraphics[margin = 0cm 0cm 0cm -1.15cm, width=0.072\textwidth,valign=c]{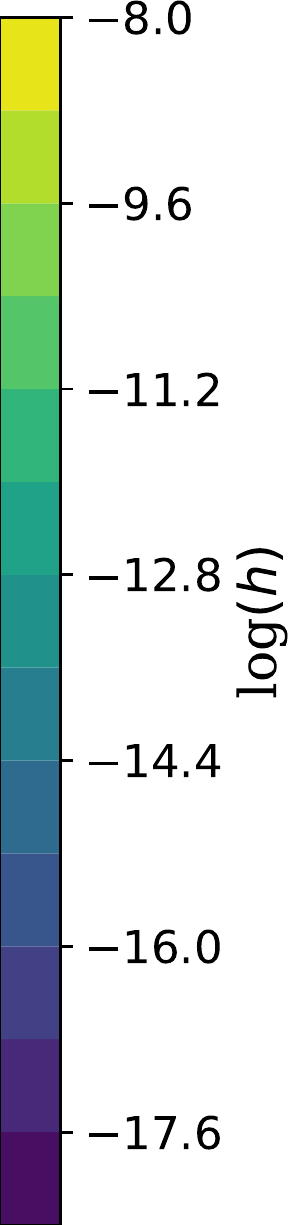}}

\caption{Strong lensing increases GW detection prospects in PTAs by $\sqrt{\mu}$ and EM detection prospects by $\mu$. High magnification also results in the amplification of the offset between sources of emission. We present the three lensing scenarios as in  \autoref{fig:param_map}, for $f_\mathrm{GW}=13$~nHz and assume a 0.1 mas angular resolution limit for ngVLA \citep{Reid2018} --- the dashed orange curve. Other GW frequencies are explored in \autoref{fig:resmaps}. 
SMBHBs with parameters in the mass-redshift space lying above e.g. the SKA-PTA and ngVLA curves will have both detectable GW and EM signals. \textit{Left Panel:} SIS lens model with $\mu=3$ and an offset amplification factor of $\sim$ 2. \textit{Center Panel:} SIE lens model with $\mu=30$ and an offset amplification factor of $\sim$ 20. \textit{Right Panel:} SIE lens model with $\mu=100$ and an offset amplification factor of $\sim$ 60.}
\label{fig:angular_res_limit}
\end{figure*}

\begin{figure*}[t]
\centering
\subfloat{\includegraphics[width=0.3075\textwidth,valign=c]{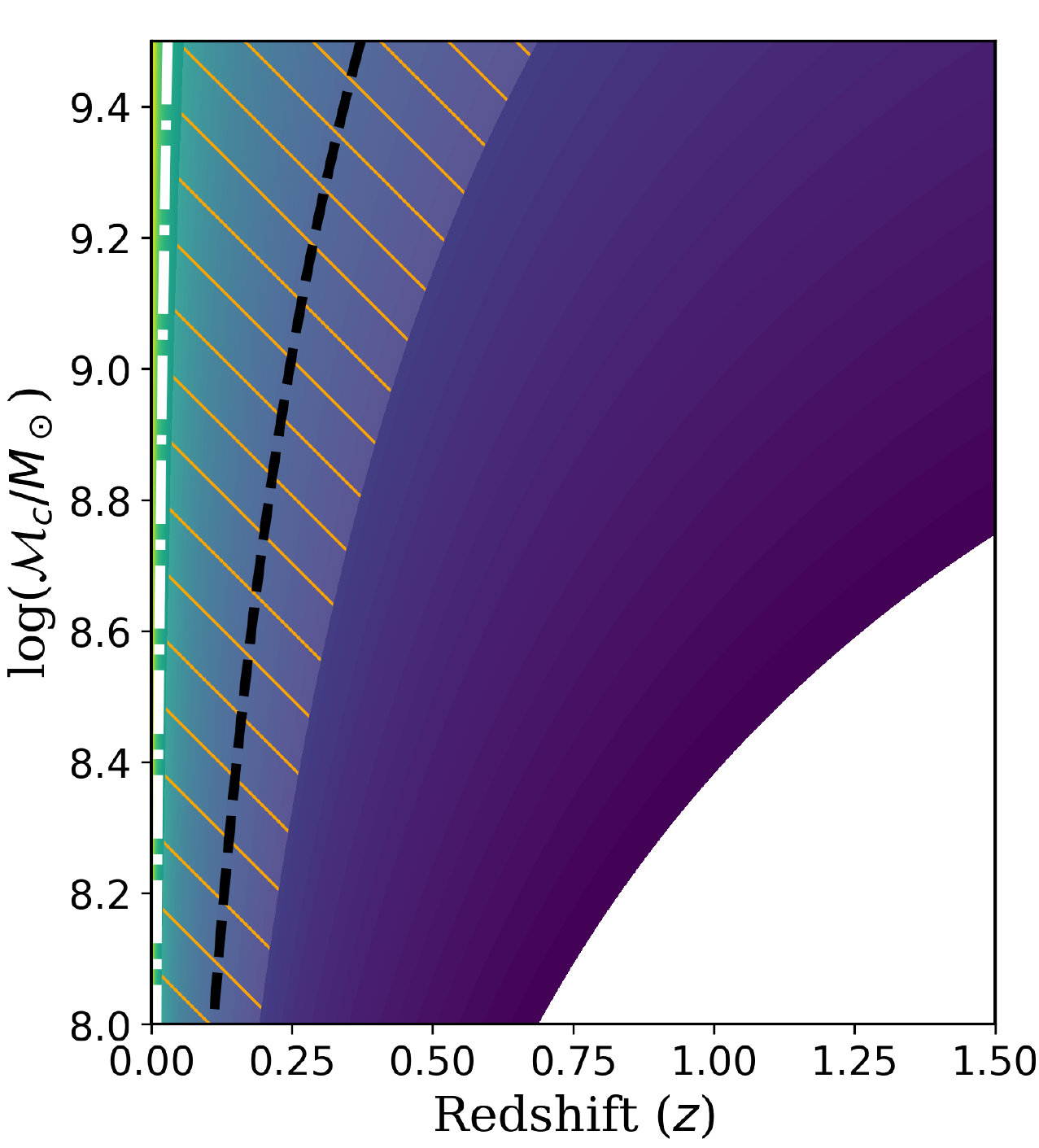}}
\hfill
\subfloat{\includegraphics[width=0.275\textwidth,valign=c]{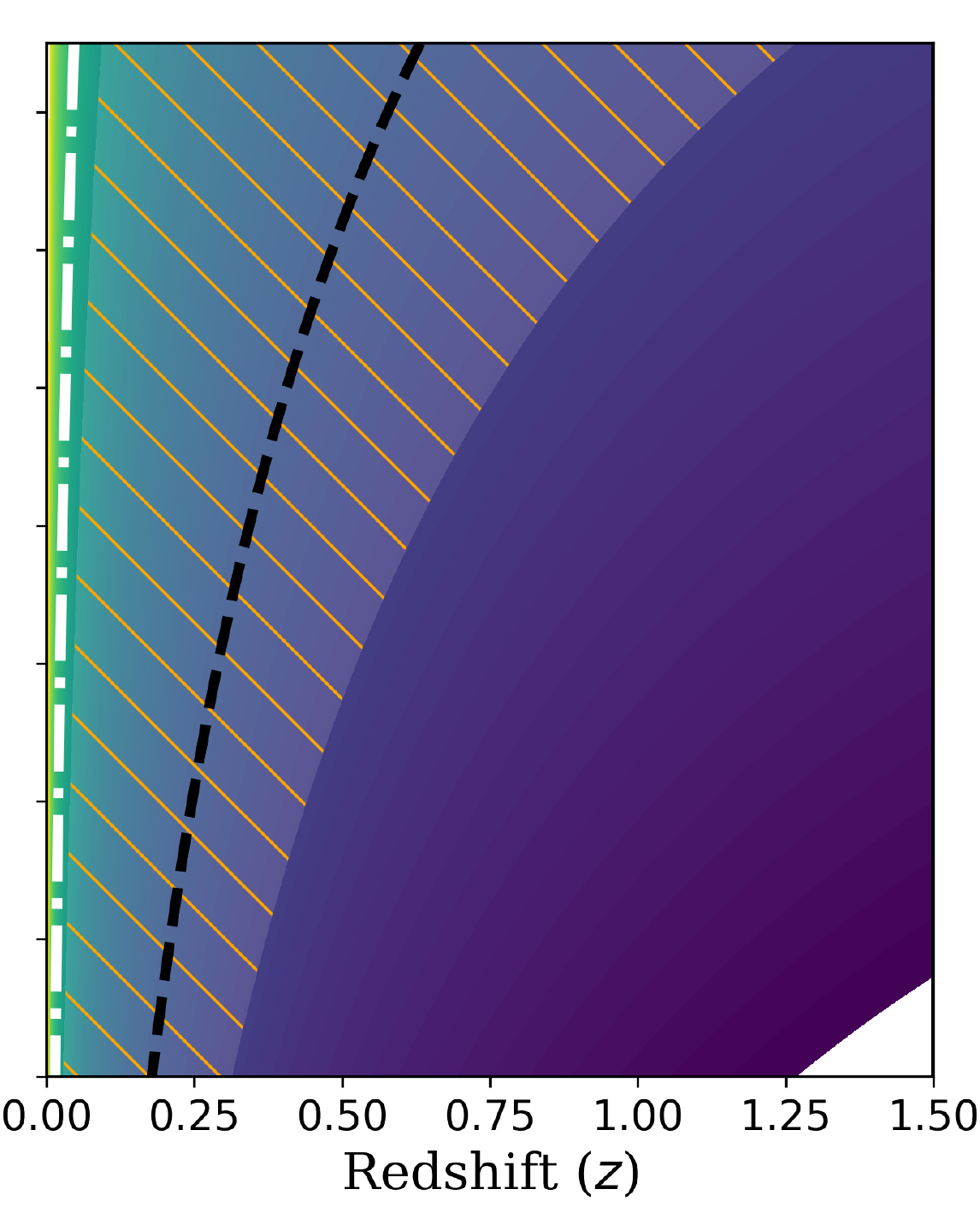}}
\hfill
\subfloat{\includegraphics[width=0.275\textwidth,valign=c]{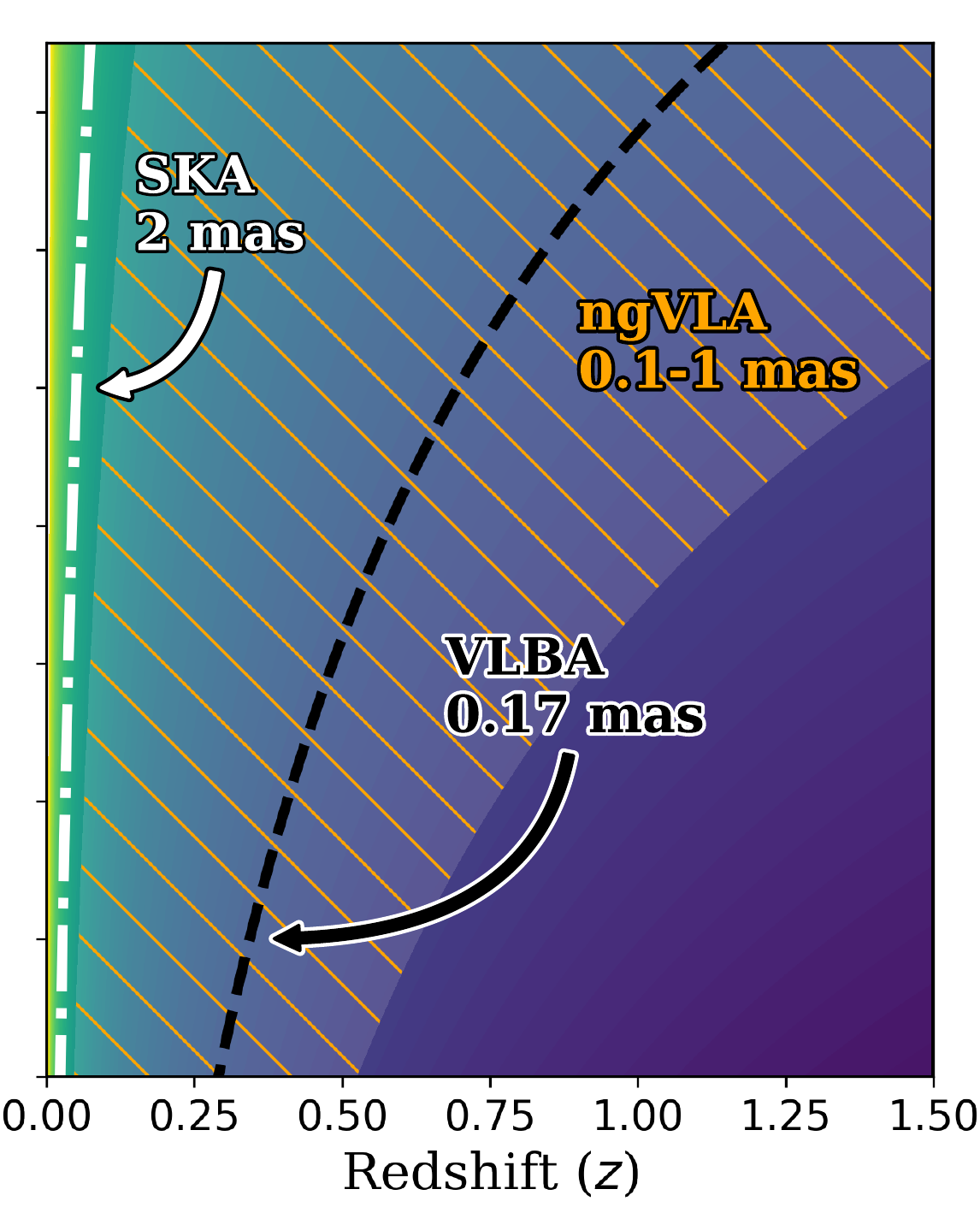}}
\hfill
\subfloat{\includegraphics[margin=0cm 0cm 0cm -1.03cm, width=0.0598\textwidth,valign=c]{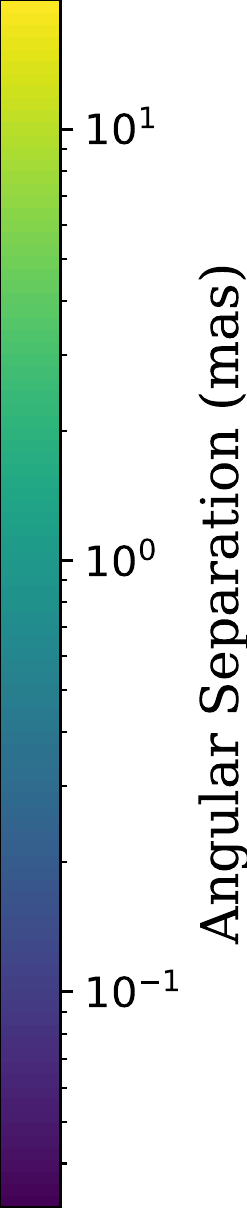}}

\caption{Lower frequency SMBHBs have wider binary separations, making their lensed images more easily detectable. For a few nearby SMBHBs with exceptionally high magnifications of, say, $\mu=100$, it may be possible to resolve the individual SMBHs, enabling simultaneous observations of gravitational and EM radiation. Binaries can be sufficiently low-frequency and high mass to be compelling strongly lensed multi-messenger systems, as in the right panel. Moreover, the flux magnification of those binaries in caustic configurations introduces magnification bias, making these sources easier to detect in the first place. Higher frequency SMBHBs are also excellent candidates for multi-messenger study due to a larger $\dot{f}$,  \autoref{fig:fdotcolormap}, however GWs from these sources are more difficult to detect \citep{NANOGrav11yr}. While ngVLA will provide the best resolution for the SMBHs, some of the lowest frequency systems can also be imaged with VLBA and SKA. \textit{Left Panel:} $f_\mathrm{GW} = 20$ nHz. \textit{Center Panel:} $f_\mathrm{GW} = 10$ nHz. \textit{Right Panel:} $f_\mathrm{GW} = 5$ nHz.}
\label{fig:resmaps}
\end{figure*}

From the EM observation perspective, the most interesting feature of strongly lensed SMBHBs is the amplified offset of the BHs within the images. High-magnification sources near lens caustics experience an order of magnitude offset amplification, greatly improving EM resolution efforts \citep{Barnacka}. To evaluate the observational prospects for strongly lensed EM counterparts of SMBHBs, we compute their angular separations as shown in \autoref{fig:Nlensed_mu}. 
For radio observations, we set an upper bound for resolvable sources with ngVLA's best expected angular resolution of $\sim$0.1 mas \citep{Reid2018}. 
We outline the overlap in ngVLA's resolution range (0.1$-$1 mas) with the detectable parameter space for GWs in \autoref{fig:angular_res_limit}. Interestingly, for the $\mu=100$ case, the ngVLA resolution limit lies above the SKA-PTA GW detection curve and below the NANOGrav 11-year curve for $0.25 \lesssim z \lesssim 1.25$. We find $ \geq 1$ strongly lensed SMBHB for $\mu=100$ as early as $z \approx 1.1$ (see \autoref{fig:Nlensed_mu}). These sources could therefore have resolvable EM counterparts detectable with the ngVLA and serve as exciting multimessenger targets only due to strong lensing.

While we present the detectable parameter space in \autoref{fig:angular_res_limit} for sources with $f_{\text{GW}} = 13$ nHz; the most feasibly detectable binaries will be low-frequency ones. As such, they will have greater separations, and, thus, produce more easily resolvable EM counterparts. For example, a lensed binary with $f_\mathrm{GW} = 1$ nHz and $\mu = 100$ is guaranteed to have resolvable EM counterparts with ngVLA. 

While higher frequency GW sources have the benefit of a greater $\dot{f}$ (see \autoref{subsec:fdot}), they are more difficult to detect. On the other hand, SMBHBs with lower $f_\mathrm{GW}$ can have their EM signals resolved at greater redshifts and at smaller chirp masses, as shown in \autoref{fig:resmaps}. In fact, the lower the frequency of the binary, the better chance we have at successfully imaging the individual SMBHs within a lensed image using multiple instruments: SKA will have an angular resolution of up to 2 mas \citep{theSKA}, and VLBA of about 0.17 mas, conservatively\footnote{\url{https://science.nrao.edu/facilities/vlba/docs/manuals/oss/referencemanual-all-pages}}. 
Furthermore, high-magnification SMBHBs, such as those whose binary separations we explore in \autoref{fig:resmaps}, must exist in caustic configurations \citep{Barnacka}. As discussed earlier, we know that lensed pair events in caustic configurations exhibit an offset amplification \citep{Barnacka}. For a lensed SMBHB with total magnification $\mu=100$, its separation is amplified by a factor of about 60 at most, depending on the pair's angle with respect to the caustic \citep{Barnacka}. This offset amplification is effectively an order of magnitude improvement to the angular resolution of the imaging instrument being used to resolve the binary, and thus it is the most valuable benefit that strong lensing contributes to resolving EM counterparts of nHz GWs.

It is important to consider observational capabilities in the case where discrete SMBHs cannot be resolved but individual images can.
For lenses modeled with an SIS, image separations are expected to be on the order of arcseconds (see \autoref{tab:time_delays}). Radio observations with SKA and ngVLA, and IR variability may be potentially detected by JWST (resolution of 0.068 arcseconds; \citealt{IR2,IR1,JWST}). We can also explore the use of planned future missions like ESA's ATHENA with an angular resolution of 5 arcseconds \citep{Athena}, to observe X-ray emissions and spectral profiles \citep{X-ray1, X-ray2}. Particularly for lower-frequency, greater separation binary sources, each BH may map to its \textit{own} family of images in the image plane. Because image configurations are very well-informed by the source positions in the source plane, we should be able to match the images to the correct corresponding BH. Then, to electromagnetically resolve the binary, we would only need to resolve one image from each BH. This approach effectively improves the angular resolution of our instruments even beyond the offset amplification, demonstrating yet another value of observing strongly lensed sources.

\subsection{Resolvability of GW Signals}
\label{subsec:fdot}
Time delays may permit PTAs to track a GW signal's evolution between lensed images of a SMBHB system. We explore four reasonable and interesting cases for an SIS lensing configuration in \autoref{tab:time_delays} to demonstrate typical expected time delays. While these are helpful example calculations to demonstrate the scaling of time delays with lens properties and position of the source, we also defer to work done by \cite{Oguri+2002} and \cite{Jana+2022} on time delay statistics, where the authors report a $\Delta t$ distribution in the SIS lens approximation that peaks around  $\sim$years for image separations of $\sim$arcseconds and for various cosmological models.

\begin{table}[h!]
\centering
\begin{tabular}{|c|c|c||c|}
    \hline
    $\beta$ (") & $\theta_E$ (") & $\sigma_v$ (km/s) & $\Delta t$ (yr) \\
    \hline\hline
    0.5 & 1.8 & 250 & 2.3 \\
    \hline
    0.5 & 2.6 & 300 & 3.3 \\
    \hline
    1 & 1.8 & 250 & 4.6 \\
    \hline
    1 & 2.6 & 300 & 6.6 \\
    \hline
\end{tabular}

\caption{PTAs may resolve multiple time-delayed copies of a strongly lensed GW signal in different PTA frequency bins. For the various SIS lensing configurations studied here, we compute the time delays using \autoref{eq:timedelay}. We choose $\sigma_v$ to model a massive elliptical galaxy \citep{sigma_v}, and we calculate $\theta_E$ directly using \autoref{eq:eradius}. All configurations assume a foreground lens position of $z=0.5$ \citep{Tewes2013}. In this case, the typical time delays are on the order of years, which is characteristic of galaxy cluster lenses.}
\label{tab:time_delays}
\end{table}

\begin{figure}[b]
    \centering
    \includegraphics[width=0.7\linewidth]{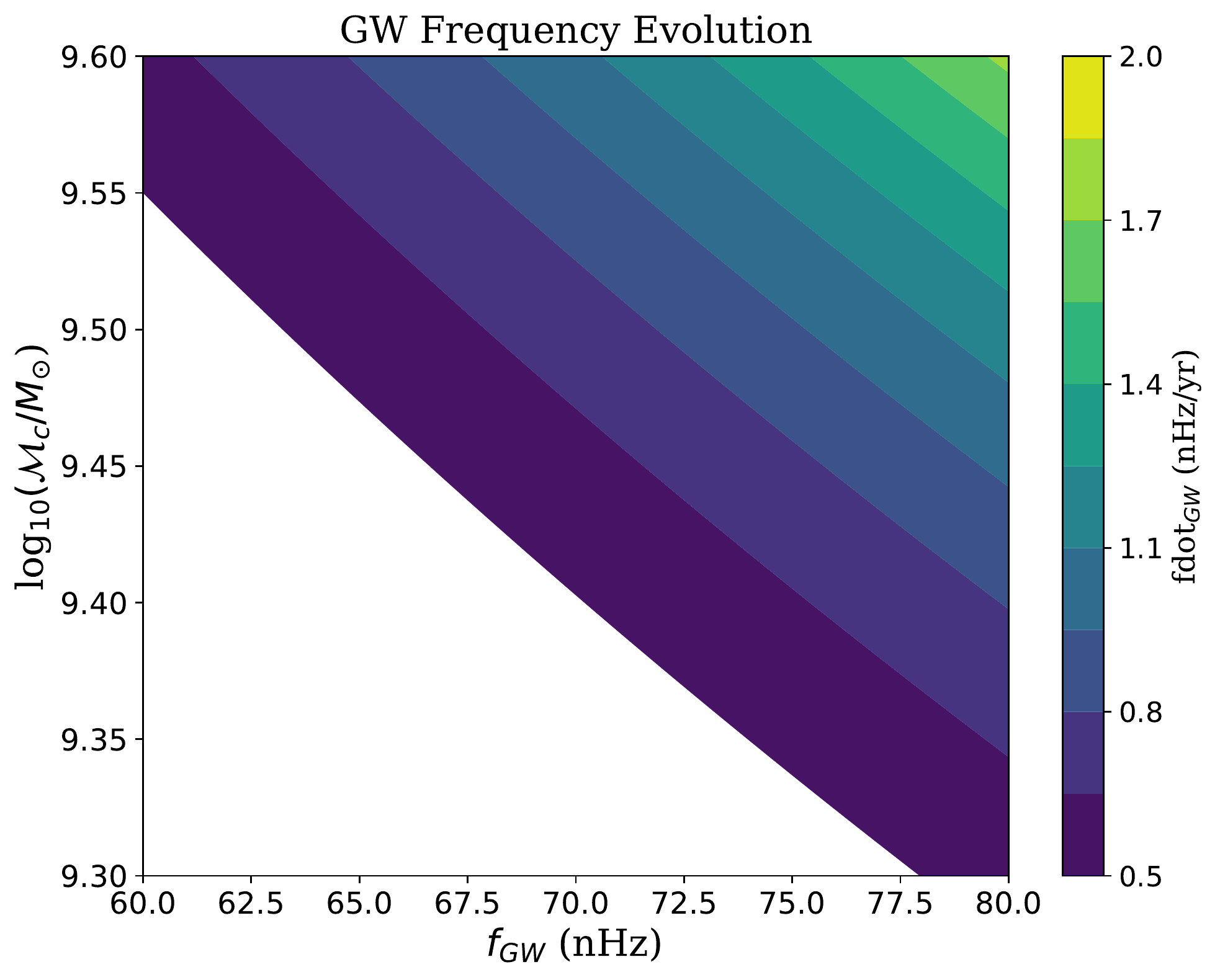}
    \caption
    {
    For a lensed SMBHB with sufficiently large $\dot{f}_\mathrm{GW}$, PTAs could detect multiple time-delayed GW signals in frequency bins, tracking the signal evolution of the system. Resolvable GW signals for frequency bin widths of $\sim$1 nHz \citep{freq_bin} and time delays of $\sim$ years warrant sources with $\dot{f} \geq 0.5$ nHz/yr. Higher frequency systems are good candidates for tracking such GW signal evolution, but they are not as easily detectable with PTAs \citep{NANOGrav11yr}.
    }
    \label{fig:fdotcolormap}
\end{figure}

Reasonable time delays are on the order of a few years, and PTAs use frequency bin widths as narrow as $\sim$1 nHz for the longest $T$. So, SMBHB systems with $\dot{f} \geq 0.5$ nHz/yr should emit lensed GW signals that are resolvable in different PTA frequency bins. \autoref{fig:fdotcolormap} shows that these systems are the most massive lensed SMBHBs emitting high-frequency GWs in the PTA band.

It is important to consider the possibility of PTAs registering the lensed images of a GW signal as an interfered signal. Theoretically, for an SIS lens model and a range of frequencies that includes the PTA band, there is equal chance that the interference would be constructive or destructive (Barnacka \& Mingarelli, in preparation). Furthermore, if it is constructive, the interfered signal would have a boosted magnification by an additional factor of 2 (Barnacka \& Mingarelli, in preparation). However, work still in preparation demonstrates that, for frequencies so low as $\sim$nHz, the effects of interference on signal amplitude are negligible (Barnacka \& Mingarelli, in preparation). We may also consider how to distinguish in the data between two lensed images of one GW signal vs. two separate signals coming from two separate sources of CWs. Work done by \cite{Haris+} demonstrates via a Bayesian inference technique that the distribution of time delays from lensed merger events peaks much lower than that of pairs of unlensed merger events.

Comparing \autoref{fig:resmaps} and \autoref{fig:fdotcolormap}, we find that EM counterpart resolvability and GW signal resolvability favor opposite ends of the PTA frequency band. In order to reconcile the two parameter spaces for a truly multi-messenger ``golden binary" system, we must find a highly magnified and hence likely SIE-lensed, massive SMBHB system in a strong caustic configuration. Such a lensed system would produce the richest data - the most optimal combinations of wide BH separations and long time delays. 

\section{Discussion}

We investigate the detection prospects for a population of strongly lensed, GW-emitting SMBHB systems out to $z=2$. Using a physically motivated $\mu$ distribution, we anticipate a detection of at least one such source. Furthermore, we can expect to detect up to an additional $\sim30$ such binaries due to strong lensing, if we use the QLF to estimate the number of SMBHBs in the most optimistic case of fixed $\mu=100$ for all sources. In fact, the steep QLF curves in \autoref{fig:Nlensed_mu} show that there exists a higher-redshift population of detectable sources that PTAs cannot probe without strong lensing; these come into view into the PTA solely due to the effects of magnification from strong lensing.

We expect PTAs to detect a population of about $\sim 800$ unlensed SMBHBs within $z \leq 2$, with an additional few SMBHBs due to lensing, see ``integrated'' result in \autoref{fig:Nlensed_mu}. The fraction of detectable lensed events is admittedly small, but also very conservative due to the fact that we adopt an SIS lens model. An SIE lens model would likely significantly increase this fraction of detected lensed events due to the much higher magnifications from caustic configurations and magnification bias. We are actively pursuing this line of research.

While we confirm that we can expect the detection of a lensed nHz GW signal in the SKA era, the most interesting such sources will be those with high magnification. We explore the extent of the multi-messenger astrophysics we can extract from particularly higher-magnification strongly lensed SMBHBs. 
The most promising solution for a ``golden binary" system, with overlapping EM and GW resolvability parameter spaces (\autoref{fig:resmaps} and \autoref{fig:fdotcolormap}) is a lensed SMBHB spanning the caustic. Caustic configurations produce the greatest magnifications and source offset amplifications, making these lensed events the most readily detectable ones. In future work, we plan to calculate the rates and observability of counter-parts using an SIE lens model explicitly.

We note that intriguing multi-messenger systems whose multiple images we can resolve, if not the individual BHs themselves, can be found. These images may wobble as the GW sources move, providing new insights into their structure and dynamics. We can resolve these strongly lensed SMBHBs with
a range of multi-wavelength EM instruments - ngVLA \citep{Reid2018}, SKA \citep{theSKA}, JWST \citep{JWST}, and possibly ATHENA \citep{Athena}. In particular, we find that the ngVLA is the ultimate instrument to extract as much multi-messenger information from such systems as we can \citep{Wrobel&Lazio}. Even if we cannot resolve the individual SMBHs within the lensed images, \cite{Wrobel&Lazio} show how ngVLA can astrometrically monitor such SMBHBs with very small separations to observe SMBH reflex motion. This opens up the discovery space to find a lensed SMBHB system that has a high enough $f_\mathrm{GW}$ to have its orbital motion astrometrically resolved with ngVLA, multiple images resolved by ngVLA, and GW signal evolution resolved by PTAs.

Our results motivate targeted searches for lensed SMBHBs now, as in \cite{3c66b}, and especially in the upcoming SKA-PTA era. We can expect to identify about 3000 strongly lensed quasars with Gaia \citep{Finet+2012}, as well as up to $\sim$10$^5$ strongly lensed systems with surveys by SKA and Euclid \citep{Laureijs+2011, SKAEuclid} in the near future, giving us our much needed targets for single source GW search efforts.

Strong gravitational lensing is a unique and promising tool that enhances PTA detection prospects. Moreover, the information we glean from SMBHBs via multiple time-delayed magnified images could be crucial in understanding SMBHB dynamics via their direct observation. Amplified images and source offsets will lend insight to emitting regions of the SMBHBs by effectively improving the resolution of our imaging tools \citep{Barnacka}. Additionally, strongly lensed SMBHBs will serve as useful cosmological probes: both the number of these events and the time delays measured between images of the source can put constraints on cosmological parameters as predicted in works like \cite{Jana+2022}.
  
Out to $z=2$, using an SIS lens model, we find at least $1$ additional SMBHB which would be detectable with PTAs entirely due to strong lensing, with at least one such system as nearby as $z \approx 1.1$ in the most optimistic case. For low-frequency binaries,
the future capabilities of ngVLA may allow us to image their discrete orbiting BHs for the first time. However, at such low frequencies the GWB must be successfully subtracted to make such a detection possible \citep{Xin}. When detected, these strongly lensed binaries, crawling towards coalescence, will be some of the richest multi-messenger systems in the Universe.

\section*{Acknowledgements}
\noindent The authors thank Ken Olum for useful discussions. This research was supported in part by the National Science Foundation under Grants No. NSF PHY-1748958, PHY-2020265, and AST-2106552. The Flatiron Institute is supported by the Simons Foundation. PN gratefully acknowledges support at the Black Hole Initiative (BHI) at Harvard as an external PI with grants from the Gordon and Betty Moore Foundation and the John Templeton Foundation.

\appendix \label{sec:Appendix}
Here we give a more detailed outline of our conversion from a QLF \citep{QLF} to the BHMF \citep{Andrew} used to compute the detectable strongly lensed SMBHB population. \cite{QLF} construct an observed bolometric QLF from a collection of observational data sets, fit with a double power-law form. The QLF also takes the form
\begin{equation}
    \phi(L) = \int \dot{\phi}(M_{\text{BH}})\frac{dt}{d\,\log L}(L|M_{\text{BH}})\,d\log M_{\text{BH}}\, ,
    \label{eq:QLFconvol}
\end{equation}
such that it is built from a convolution of the quasar, or BH, formation rate, $\dot{\phi}(M_{\text{BH}})$, and differential lifetime, $\frac{dt}{d\,\log L}$. The theoretical light curve model used to determine the differential quasar lifetime is derived from BH-growth-driven merger simulations \citep{Hopkins+2006}. To build up a BH population, \cite{QLF} de-convolve the QLF with this model to isolate and fit the BH formation rate similarly to \autoref{eq:QLF}:
\begin{equation}
    \dot{\phi}(M_{\text{BH}}) = \frac{\dot{\phi}_*}{(M_{\text{BH}}/M_*)^{\eta_1} + (M_{\text{BH}}/M_*)^{\eta_2}}\, ,
    \label{eq:formation_rate}
\end{equation}
with normalization $\dot{\phi}_*$, break BH mass $M_*$, and faint and bright-end slopes $\eta_1$ and $\eta_2$, respectively. We follow the prescription laid out in \cite{Andrew} and multiply by $dt/dz$ to write the BH formation rate as a smooth BH mass function:
\begin{equation}
    \phi(M_{\text{BH}}, z) = \dot{\phi}(M_{\text{BH}})\frac{dt}{dz} = \frac{d\Phi_\text{BH}}{d\log M_{\text{BH}}\,dt}\frac{dt}{dz}\, .
\end{equation}

\begin{figure}[h!]
    \centering
    \includegraphics[width=0.50\textwidth]{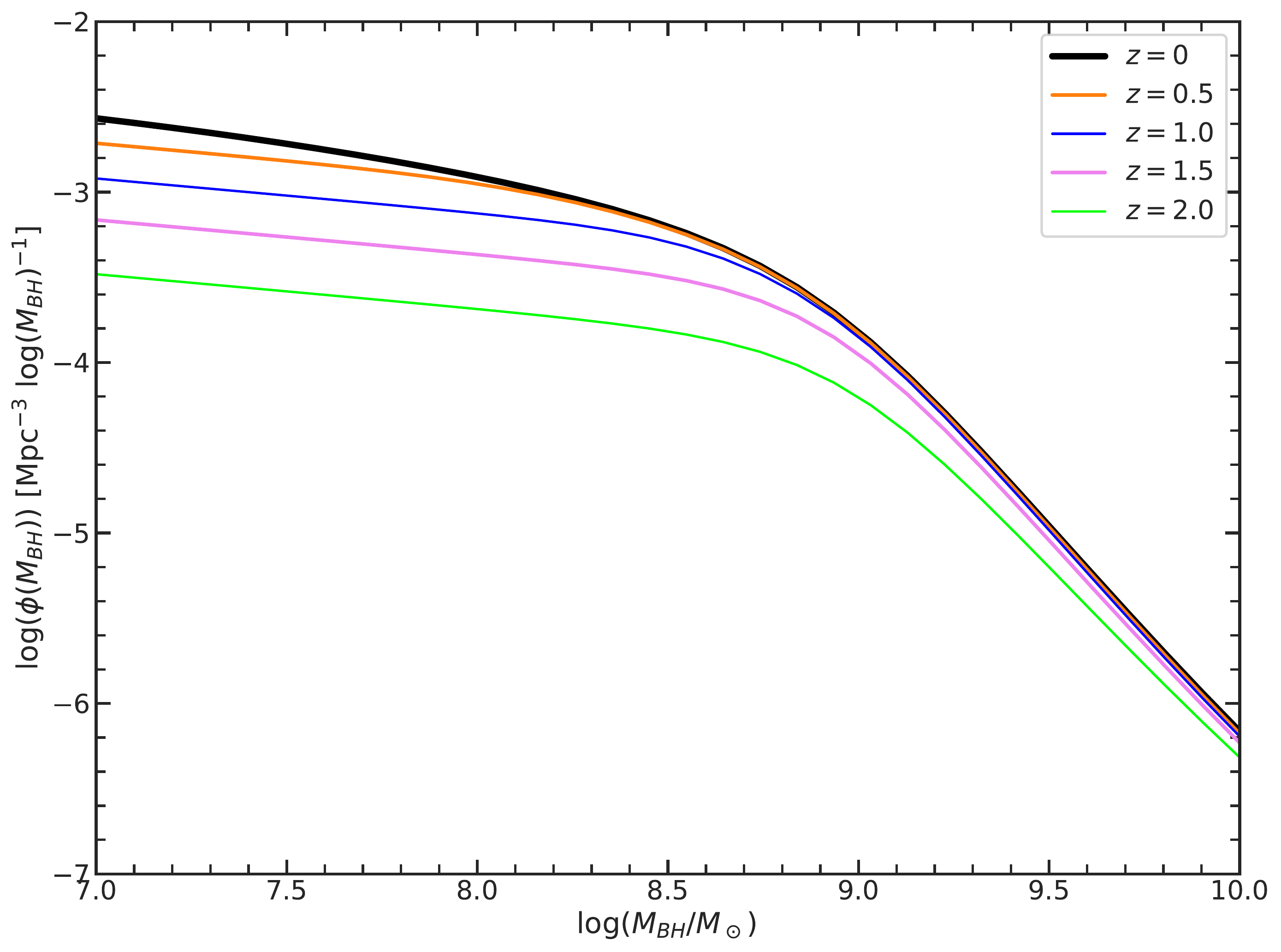}
    \includegraphics[width=0.49\textwidth]{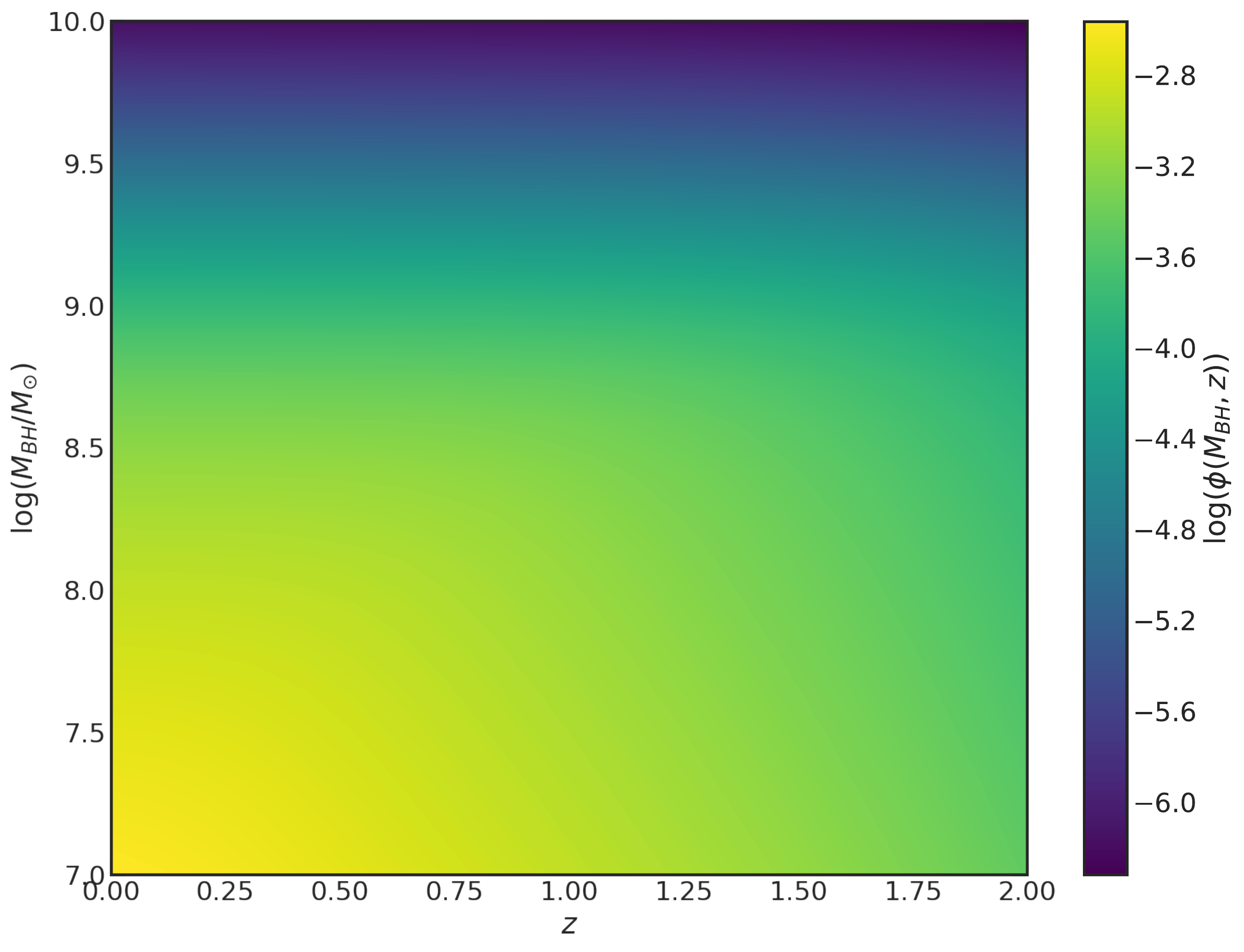}
    \caption{\textit{Left:}The log-log plot of the integrated BHMF against the BH mass, calculated at various redshifts. The $z=$ 0, 1, 2 curves can be compared to and confirmed to match those from the bottom center panel of Fig. 10 in \cite{QLF}. \textit{Right:} 2D plot of the BHMF after converting the BH formation rate into a smooth function of $z$.}
    \label{fig:QLF_demo}
\end{figure}

In \autoref{fig:QLF_demo}, we demonstrate that our smoothly $z$-evolving BHMF still matches that of \cite{QLF} when marginalized over redshift. We also present a 2-dimensional representation of how the distribution of BHs evolve in both $z$ and BH mass. The change in the distribution across $z$ becomes minimal for BH masses greater than $\sim 10^9 M_\odot$. 

\end{document}